\documentclass[aps,pra,superscriptaddress,reprint,a4paper,nofootinbib]{revtex4-2}

\usepackage[utf8]{inputenc}
\usepackage[T1]{fontenc}
\usepackage{amsmath,amsfonts,amssymb}
\usepackage{mathtools} 
\usepackage{graphicx}
\usepackage{xcolor}
\usepackage{comment}
\usepackage{tikz} 
\usepackage{gensymb}

\usepackage{bm}

\DeclareMathAlphabet{\mathbbb}{U}{bbold}{m}{n}

\usepackage[normalem]{ulem}

\usepackage{xparse} 

\graphicspath{ {./figs/} }
\DeclareGraphicsExtensions{.pdf,.png}
\usepackage{siunitx}

\definecolor{myurlcolor}{rgb}{0,0,0.7}
\definecolor{myrefcolor}{rgb}{0.8,0,0}
\usepackage{hyperref}
\hypersetup{
    unicode=true, bookmarksnumbered=false, bookmarksopen=false, breaklinks=false, pdfborder={0 0 0},
    colorlinks=true,
    linkcolor=myrefcolor,citecolor=myurlcolor,urlcolor=myurlcolor
}

\newcounter{repeatedfootnote}
\newcommand{\myfootnote}[1]{%
    \footnote{#1}%
    \setcounter{repeatedfootnote}{\value{footnote}}%
}

\newcommand{\0}{\mathbbb{0}}
\newcommand{\1}{\mathbbb{1}}
\newcommand{\TT}{\mathsf{T}}

\newcommand{\trm}[1]{\textrm{#1}}

\newcommand{\mrm}[1]{\mathrm{#1}}

\newcommand{\eref}[1]{(\ref{#1})}

\newcommand{\eqnref}[1]{Eq.~(\ref{#1})}
\newcommand{\eqnsref}[2]{Eqs.~(\ref{#1}-\ref{#2})}
\newcommand{\figref}[1]{Fig.~\ref{#1}}
\newcommand{\tabref}[1]{Table~\ref{#1}}
\newcommand{\secref}[1]{Sec.~\ref{#1}}
\newcommand{\appref}[1]{App.~\ref{#1}}

\newcommand{\citeref}[1]{Ref.~\cite{#1}}
\newcommand{\citerefs}[1]{Refs.~\cite{#1}}

\renewcommand{\vec}[1]{\bm{#1}}
\newcommand{\mat}[1]{\mathbf{#1}}

\newcommand{\x}[1]{\vec{x}_{#1}}

\newcommand{\y}[1]{\vec{y}_{#1}} 
\newcommand{\Y}[1]{\vec{Y}_{#1}} 

\newcommand{\E}{\mathbb{E}}
\newcommand{\EE}[2]{\E_{#1}\!\left[#2\right]}

\newcommand{\est}[1]{\tilde{#1}}
\newcommand{\param}{\theta}
\newcommand{\paramV}{\vec{\param}}
\newcommand{\estx}[1]{\est{\vec{x}}_{#1}}

\newcommand{\errormat}{\mat{\Sigma}}

\newcommand{\BI}{I_\mrm{B}}
\newcommand{\FI}{I_\mrm{F}}
\newcommand{\BIM}{\mat{I}_\mrm{B}}
\newcommand{\FIM}{\mat{I}_\mrm{F}}

\newcommand{\cL}{\mathcal{L}}
\newcommand{\cN}{\mathcal{N}}

\newcommand{\cJ}{\mathcal{J}}

\newcommand{\Tr}{\mathrm{Tr}}
\NewDocumentCommand\trace{g}{
  \IfNoValueTF{#1}
    {\Tr}
    {\Tr\!\left\{#1\right\}}
}
\newcommand{\tr}[1]{\trace{#1}}

\newcommand{\larmor}{\omega}

\begin{document}

\title{Optimal and efficient inference tools for field tracking with precessing spins}

\newcommand{\orcidKD}{0000-0002-5460-5756}
\newcommand{\orcidPB}{0000-0003-4496-8673}
\newcommand{\orcidJK}{0000-0001-8211-0016}
\newcommand{\orcidDMA}{0000-0003-4451-2203}
\newcommand{\orcidAS}{0000-0001-8731-9513}
\newcommand{\orcidMWM}{0000-0001-8949-9407}

\newcommand{\orcidauthorA}{\orcidKD}
\newcommand{\orcidauthorB}{\orcidPB}
\newcommand{\orcidauthorC}{\orcidDMA}
\newcommand{\orcidauthorD}{\orcidAS}
\newcommand{\orcidauthorE}{\orcidMWM}
\newcommand{\orcidauthorF}{\orcidJK}

\definecolor{lime}{HTML}{A6CE39}
\DeclareRobustCommand{\orcidicon}{
    \begin{tikzpicture}
    \draw[lime, fill=lime] (0,0) 
    circle [radius=0.16] 
    node[white] {{\fontfamily{qag}\selectfont \tiny ID}};
    \draw[white, fill=white] (-0.0625,0.095) 
    circle [radius=0.007];
    \end{tikzpicture}
    \hspace{-2mm}
}
\foreach \x in {A, ..., Z}{\expandafter\xdef\csname orcid\x\endcsname{\noexpand\href{https://orcid.org/\csname orcidauthor\x\endcsname}
            {\noexpand\orcidicon}}
}

\newcommand{\ICFO}{ICFO - Institut de Ci\`encies Fot\`oniques, The Barcelona Institute of Science and Technology, 08860 Castelldefels (Barcelona), Spain}
\newcommand{\ICREA}{ICREA - Instituci\'{o} Catalana de Recerca i Estudis Avan{\c{c}}ats, 08010 Barcelona, Spain}
\newcommand{\CENT}{Centre of New Technologies, University of Warsaw, Banacha 2c, 02-097 Warsaw, Poland.}
\newcommand{\IFPAN}{Institute of Physics, Polish Academy of Sciences, Aleja Lotnik\'{o}w 32/46, 02-668 Warsaw, Poland.}
\newcommand{\AGH}{AGH University of Krak\'{o}w, Faculty of Electrical Engineering, Automation, Computer Science and Biomedical Engineering, Department of Automation and Robotics, al.~A.~Mickiewicza 30, 30-059 Krak\'{o}w, Poland.}

\author{Klaudia Dilcher\orcidA}
\email{k.dilcher@cent.edu.pl}
\affiliation{\CENT}
\author{Piotr Bania\orcidB}
\email{pba@agh.edu.pl}
\affiliation{\AGH}
\affiliation{\IFPAN}
\author{Diana Méndez-Avalos\orcidC}
\affiliation{\ICFO}
\author{Aleksandra Sierant\orcidD}
\affiliation{\ICFO}
\author{Morgan W. Mitchell\orcidE}
\affiliation{\ICFO}
\affiliation{\ICREA}
\author{Jan Ko\l{}ody\'{n}ski\orcidF}
\email{jankolo@ifpan.edu.pl}
\affiliation{\IFPAN}
\affiliation{\CENT}

\date{\today}

\begin{abstract}
Precise, real-time monitoring of magnetic field evolution is important in applications including magnetic navigation and searches for physics beyond the standard model. One main field-monitoring technique, the spin-precession magnetometer (SPM), observes electron, nucleus, color center, or muon spins as they precess in response to their local magnetic field. Here, we study Bayesian signal-recovery methods for SPMs in the free-induction decay (FID) mode. In particular, we study tracking of field changes well within the coherence time of the spin system, and thus well beyond the response bandwidth, as in [R.~Jim\'{e}nez-Mart\'{i}nez~\emph{et~al.~}\href{https://doi.org/10.1103/PhysRevLett.120.040503}{Phys.~Rev.~Lett.~\textbf{120}, 040503 (2018)}].  We derive the Bayesian Cram\'er-Rao bound that dictates the ultimate precision in estimating the Larmor frequency, which we show to be attained by the computationally expensive prediction error method (PEM). Relative to this benchmark, we show that the extended Kalman filter (EKF) and cubature Kalman filter (CKF) offer  near-optimal tracking that is also computationally efficient, with the use of the latter giving better results only for a large spin number. Focusing thus on the EKF, we show that it is sufficient to accurately track fluctuating and unknown transient signals. Our methods can be easily adapted to other types of sensors undergoing nonlinear dissipative dynamics and experiencing intrinsic Gaussian-like stochastic noises.
\end{abstract}

\maketitle

\section{Introduction}
\label{sec:intro}
Spin-based quantum sensors~\cite{Jackson2023,Aslam2023} use the precession of their spin degree of freedom to precisely sense external magnetic fields and their variations. They provide an important alternative to superconducting quantum interference devices (SQUIDs)~\cite{SQUIDS}, being able to reach subfemtotesla sensitivities~\cite{ATOMICkominis2003subfemtotesla} without the need for cryogenic cooling. The two promising platforms for these \emph{spin-precession magnetometers} (SPMs) are hot-vapour atomic ensembles~\cite{Budker2007} and colour centres in crystals~\cite{taylor2008high, casola2018probing}. Both are well suited for real-time sensing tasks---a strength of the latter is the ultra-high spatial resolution, whereas the former offers the highest sensitivities \cite{DangAPL2010} being fundamentally limited by intrinsic noises that produce relaxation and associated fluctuations of the ensemble spin~\cite{BudkerKimball}. These features enable emerging technologies such as magnetic navigation in GPS-denied environments~\cite{Canciani2020,Canciani2022}, ultra-low-field NMR~\cite{Savukov2005} and MRI~\cite{Savukov2009}, and materials characterisation~\cite{Romalis2011}, as well as medical applications for sensing brain (MEG~\cite{Brookes2022}) or heart (MCG~\cite{Jensen2018, Kim2019}) signals.

The stochastic character of spin dynamics and the light-detection process in SPMs makes it natural to use signal-processing tools that operate \emph{recursively}~\cite{Ljung1987} so that the detected signal may be interpreted in a sequential manner as it is gradually collected. Such methods relying on \emph{Bayesian inference}~\cite{Sarkka,Trees1968} are then naturally suited to estimate magnetic fields that vary in time, with fluctuations of the input field being easily incorporable into the framework. Moreover, their performance can be benchmarked by deriving limits imposed on the minimal \emph{mean squared error} (MSE) applicable to any estimator~\cite{Trees2007}.

In this work, we consider a stochastic dynamical model of the SPM~\cite{Jimenez2018} operated in the free-induction decay (FID) mode~\cite{Gemmel2010,Grujic2015,Hunter2018,Yi2024,Kong2025} that correctly reproduces its measured noise spectrum for different atomic densities~\cite{Lucivero2016}, so that its optimal working conditions can be identified~\cite{Troullinou2023}. Based on the model, we derive the so-called \emph{Bayesian Cram\'{e}r-Rao Bound} (BCRB)~\cite{Trees2007} that sets the limit on the error in estimating the Larmor frequency, which cannot be surpassed by any estimator built on the data measured%
\myfootnote{
    Given any prior distribution that represents our knowledge about the field at the initial time $t=0$ with vanishing tails~\cite{Gill1995}.
}. We then demonstrate that, for any probing time and atomic density, the BCRB can be attained by resorting to the \emph{Prediction Error Method} (PEM)~\cite{Astrom1979}. The so-constructed estimator of the Larmor frequency, however, must be recomputed each time a new measurement is recorded using the full history of measurement data---making such an approach intractable in practice. 

Hence, bearing in mind practical applications that require the magnetic field to be estimated in real time, we propose Bayesian filtering techniques as a natural solution~\cite{Sarkka}. These, by relying on the Gaussian approximation, provide estimators that are computed based only on the current output of the measurement and the most recent estimates of the parameters. Their most common example is the \emph{Kalman Filter}~\cite{Kalman1960, Kalman1961} that is guaranteed to be optimal, i.e., achieve the BCRB~\cite{Trees2007}, as long as the dynamics of the estimated parameters is described by a \emph{linear} stochastic differential equation (SDE) with Gaussian noise. Unfortunately, this is not the case for any frequency estimation task that requires inference of the amplitude of an oscillatory signal---in particular, in spin-based quantum sensors in which both the precession speed and the magnitude of the spin must be estimated in real time---as any SDE describing such dynamics involves products of these two degrees of freedom.

For this reason, we resort here to \emph{nonlinear} Gaussian filters~\cite{Sarkka}, starting from the \emph{Extended Kalman Filter} (EKF)~\cite{smith1962application} that effectively performs a time-local linearisation of the dynamics when updating the estimates after recording each measurement output. Although the EKF has been known for decades, it is still the state-of-the-art method applied in many engineering disciplines and is a subject of active research in navigation systems, imaging, and many more areas~\cite{Hide2003, Jo2015, Grewal2010}. 

We then construct the \emph{Cubature Kalman Filter} (CKF)~\cite{Arasaratnam2009}, which generalises the EKF beyond linearity by relying on moment matching within the Gaussian approximation with (spherical cubature) integrals evaluated numerically up to the third order~\cite{Sarkka}. Although this requires higher computational complexity, the algorithm remains recursive and, hence, in principle tractable in real time.

In order to establish the capabilities of the EKF and CKF, we simulate the SPM scenario of~\cite{Jimenez2018} while varying the relevant experimental parameters:~probing time $t$, number of atoms involved $N$, and the measurement sampling period $\Delta$. Firstly, we consider the scenario of (i):~estimating a constant Larmor frequency, $\larmor$, which is induced by an invariant, strong but unknown magnetic field. This allows us to show that for typical $\larmor\simeq 2\pi \times \SI{10}{\kilo\hertz}$ , both the EKF and CKF reach sub-$\SI{0.01}{\hertz}$ sensitivities within the SPM coherence time, with the implementation of the CKF being essential only for large atomic ensembles with $N\gtrsim10^{11}$. Moreover, the finite sampling period does not affect the performance as long as it is less than approximately $\SI{10}{\micro\second}$. 

That is why, when moving on to tasks of sensing time-varying fields, we focus on the EKF, which we show to be efficient in (ii):~tracking fluctuating signals---we choose an Ornstein-Uhlenbeck process~\cite{gardiner2004handbook} with a stabilisation time $\approx\SI{1}{\second}$ and stochastic jumps yielding a root mean square (RMS) deviation of 1\%--10\% of $\larmor$ over the SPM's coherence time. In order to further prove the versatility of the EKF, we set it to expect a Wiener process~\cite{gardiner2004handbook} with moderate noise and show that it remains efficient in (iii):~tracking unknown transient signals---we consider overlain oscillations of $\larmor$ at approximately $\SI{0.5}{\kilo\hertz}$ and sudden jumps (step functions), both leading to variations in $\larmor$ of about 10\% ($\approx\SI{1}{\kilo\hertz}$).

The manuscript is organised as follows. In \secref{sec:model}, we describe the operation of the SPM under consideration, including the experimental parameters used in~\citeref{Jimenez2018}. In \secref{sec:infer}, we introduce the inference tools used to interpret the magnetometer output signal. In particular, we start by discussing the BCRB and the PEM algorithm in Secs.~\ref{sec:BCRB} and \ref{sec:PEM}, respectively, before introducing (nonlinear) Gaussian filters in \secref{sub:nonlinear_filters}, of which EKF and CKF are examples. Our simulations of the SPM with applications of the developed inference tools are then presented in \secref{sec:results}. However, we also provide a software library~\cite{Github}, which contains all the methods implemented. We conclude the manuscript in \secref{sec:conclusions}.

\section{Magnetometer model}
\label{sec:model}
The specific SPM we study is an optically pumped magnetometer~\cite{FabricantNJP2023} operating in the FID mode~\cite{Gemmel2010,Grujic2015,Hunter2018,Yi2024,Kong2025}:~the atomic spins are first polarised by optical pumping and then allowed to precess about the magnetic field, while also relaxing and experiencing noise processes. Other kinds of SPMs have similar dynamical descriptions, e.g., those in \citeref{Troullinou2021}. The spin system we consider is an ensemble of $N$ atoms with ground-state spin $F$. The collective spin of the ensemble is described by its total spin operator $\hat{J}$ with components $\hat{J}_\alpha$, $\alpha \in \{x,y,z\}$, and their mean values ${J}_\alpha(t)\coloneqq\Tr\{\rho(t)\hat{J}_\alpha\}$, where $\rho(t)$ is the state of the ensemble, which evolves according to a Lindblad master equation made stochastic by atomic noise processes, including scattering of probe light and diffusion of atoms into and out of the probed region~\cite{ManzanoARX2020}.

\begin{figure}[t]
    \begin{center}
      \includegraphics[width=\columnwidth]{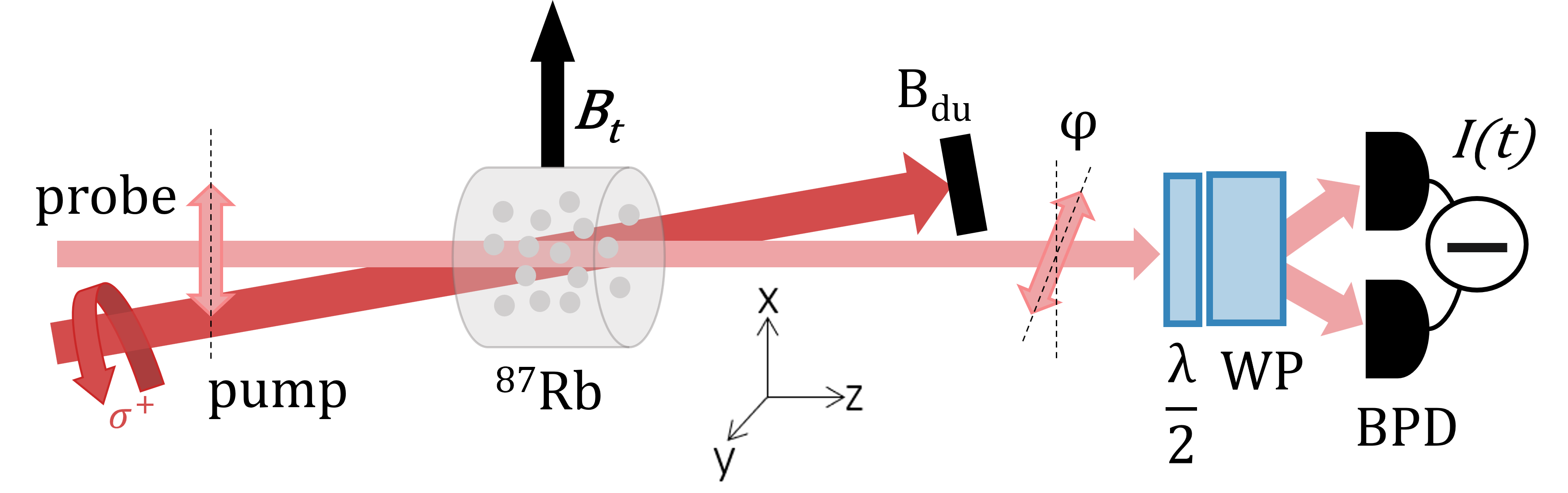}
    \end{center}
    \caption{\textbf{Setup of the simulated spin-precession magnetometer (SPM)}. The setup consists of a 3-cm-long glass cell with isotopically enriched $^{87}$Rb and 100 Torr of N$_2$ buffer gas at $\SI{100}{\degreeCelsius}$. Through the cell pass a linearly polarised probe light and a circularly polarised pump light. The rotation in the polarisation angle of the probe light after the atoms is given by $\varphi\propto\vec{J}_t\cdot \vec{e}_z$. $I(t)\propto\varphi$ denotes the photocurrent measured at time $t$, which then evolves according to \eqnref{eq:discrete_meas}. $\trm{B}_\trm{du}$---beam dumper, $\frac{\lambda}{2}$---half-wave plate, WP---Wollaston prism, BPD---balanced photodetector.
    }
  \label{fig:setup}
\end{figure}

During the preparation stage, the ensemble is pumped by a circularly polarised pump beam, resonant with an allowed transition and propagating along the $+z$ direction, as shown in \figref{fig:setup}. We assume that this pumping is complete, such that $J_z(0)= N F$, $J_y(0)=J_x(0)=0$, at $t=0$.  The pump is then suddenly switched off and the spins evolve in the presence of a magnetic field  $\vec{B}_t=B_t\vec{e}_x$ along the $x$ axis (see \figref{fig:setup}). Because of this field, the ensemble angular momentum experiences Larmor precession about the direction of $\vec{B}_t$ at the (angular) \emph{Larmor frequency}, $\larmor(t)$, which is proportional to the magnetic field:~$B_t=\larmor(t)/\gamma$, where $\gamma$ is the gyromagnetic ratio. Simultaneously, the atoms undergo isotropic relaxation that leads to an exponential decay of each $J_\alpha(t)$ at a constant rate, as well as intrinsic noise arising from optical depolarization, collisional spin randomization, and diffusion of atoms out of and into the probed region. The latter is modelled by independent Gaussian white noise in each $J_\alpha$-component. 

The precession couples the  $J_y$ and $J_z$ components, whereas after optical pumping, $J_x$ is not coupled to any other spin degree of freedom. Because of this, we can ignore $J_x$ in what follows. We define a vector denoting the collective atomic-spin components in the $y$-$z$ plane, $\vec{J}_t\coloneqq[J_y(t)\,J_z(t)]^\TT$, which is perpendicular to the $\vec{B}_t$ field (see \figref{fig:setup}). We describe its dynamics with the following \^{I}to stochastic differential equation~\cite{Jimenez2018}: 
\begin{equation}
    d\vec{J}_t = \mat{A}(t)\vec{J}_t\,dt + \sqrt{\mat{Q}}\,d\vec{w}^{(J)}_t,\\
    \label{eq:spin_dynamics}
\end{equation}
where 
\begin{equation}
    \label{eq:A_Q_matrices}
    \mat{A}(t) =    
        \begin{bmatrix}
            -\frac{1}{T_2}&\larmor(t) \\
            -\larmor(t)&-\frac{1}{T_2}
        \end{bmatrix},
    \quad
    \mat{Q} =    
        \begin{bmatrix}
            Q_y&0 \\
            0&Q_z
        \end{bmatrix},
\end{equation}
and by $T_2$ we denote the effective transverse coherence time of the atomic ensemble, while  $d\vec{w}_t^{(J)}=[dw^{(y)}_t\,dw^{(z)}_t]^\TT$ is the vector of stochastic increments representing uncorrelated zero-mean Gaussian noise in each of the spin components. Throughout this work, by $dw^{(\alpha)}_t$ we denote independent Wiener increments that satisfy $\E[dw^{(\alpha)}_t dw^{(\beta)}_{t'}]=\delta_{\alpha\beta}\delta(t-t')\, dt$ and by $\E[\dots]$ the average over stochastic fluctuations~\cite{gardiner2004handbook}. 

In the regime where probe-induced backaction can be neglected, the dominant microscopic relaxation processes are isotropic in the plane perpendicular to the magnetic field. As a result, the intrinsic spin noise is assumed to be cylindrically symmetric around the direction of $\vec{B}_t$, and we may set $Q_y=Q_z\eqqcolon Q$ in \eqnref{eq:A_Q_matrices} as the strength of atomic fluctuations. $Q$ scales as $N/T_2$ and describes the injection of uncorrelated noise at a rate $1/T_2$~\cite{Pezze_Quantum_Metrology_RevModPhys.90.035005}.

In particular, the fluctuation-dissipation theorem~\cite{gardiner2004handbook} imposes $2\mat{P}_\trm{ss}/T_2=\mat{Q}$ for the dynamics \eref{eq:spin_dynamics}, where $[\mat{P}_\trm{ss}]_{\alpha\beta}=\E[J_\alpha J_\beta]-\E[J_\alpha]\E[J_\beta]$ is the steady-state covariance of the atomic spin. Hence, for uncorrelated atoms, for which one expects $\mat{P}_\trm{ss}=(qN/2)\1_2$ with $q = F(F+1)/3$ being  the thermal state variance~\cite{MouloudakisPRA2022}%
\footnote{
    Throughout, by $\0_{d_1\times d_2}$ and $\1_d$, we denote zero and identity matrices of dimensions $d_1\times d_2$ and $d$, respectively. 
}, we may parametrise the overall atomic-noise strength as
\begin{equation}
    Q \coloneqq \frac{q\,N}{T_2(N)}.
    \label{eq:Q_eff}
\end{equation}
In what follows, we take $F=1/2$ and, hence, $q=1/4$ for simplicity.

However, as emphasised in \eqnref{eq:Q_eff}, we have indicated explicitly that $T_2$, the coherence time of the ensemble, depends on $N$, the atom number. In what follows, we use the model~\cite{Lucivero2016}:
\begin{equation}
    \frac{1}{T_2}=\Gamma+\alpha N,
    \label{eq:T_2}
\end{equation}
where $\Gamma=\Gamma_0+\beta P$ is the effective linewidth, incorporating the ``power broadening''  contribution $\beta P$ (assumed constant here), due to scattering of probe light, as described below. The term $\alpha N$ accounts for the broadening due to spin-exchange collisions~\cite{Savukov2005a}, and increases linearly with the number of atoms, given a constant $\alpha$~\cite{Lucivero2016,Delpy2023}. Hence, for high atomic densities such that $\alpha N\gg \Gamma$ in \eqnref{eq:T_2}, it follows that the atomic-noise strength \eref{eq:Q_eff} scales quadratically with the number of atoms, i.e., $Q\propto N^2$.

As illustrated in \figref{fig:setup}, the atomic ensemble is continuously probed by a separate linearly polarised optical beam that is far-detuned from the relevant atomic transition. The probe polarization experiences linear Faraday rotation by an angle proportional to $J_z$, the collective spin component aligned with the probe's propagation direction~\cite{Happer1967}. In particular, the photocurrent $I(t)$ measured by a polarimeter, such as the one depicted in \figref{fig:setup}, is proportional to $J_z(t)$ and undergoes dynamics described by another stochastic differential equation~\cite{Jimenez2018}:
\begin{equation}	
	I(t) dt = \vec{H} \cdot \vec{J}_t\,dt + dv^{(I)}_t.
    \label{eq:meas_signal}
\end{equation}
\eqnref{eq:meas_signal} is written as a function of the whole vector $\vec{J}_t$ appearing in \eqnref{eq:spin_dynamics}, however, with $\vec{H}\coloneq[0\,g_D]^\TT$ selecting only the $z$ component. The measurement strength $g_D$ is a constant determined by a particular experimental setup, i.e., the exact detuning and power of the probe beam, the optical depth of the ensemble, as well as the properties of the detector, e.g., the photodiode responsivity~\cite{Jimenez2018}. 

Similar to the atomic-noise terms in \eqnref{eq:spin_dynamics}, the increment $dv^{(I)}_t$ in \eqnref{eq:meas_signal} represents zero-mean Gaussian fluctuations, i.e., $dv^{(I)}_t \sim \cN(0, R\,dt)$, that are independent at different times, i.e., $\E[dv^{(I)}_s dv^{(I)}_t]= R\,\delta(s-t)dt$ where $R$ is now the (scalar) measurement-noise covariance. Physically, these describe the photon shot noise arising during the light-detection process, which is uncorrelated with the atomic noise, i.e., $\forall{\alpha;t,s\ge0}:\,\E[dw_t^{(\alpha)} dv^{(I)}_s] = 0$, due to the lack of measurement backaction being assumed.

In practice, however, the detection can be performed only at a finite probing rate $\Delta$, i.e., at times $t_k=k\,\Delta$ with $k\in\mathbb{N}^+$, so that to describe real-life experiments \eqnref{eq:meas_signal} should be rewritten as
\begin{equation}
    I(t_k) =
    \int_{t_k-\Delta}^{t_k}\!\!\! \mathcal{R}_k(\tau) \, \vec{H} \cdot \vec{J}_\tau d\tau + v_k^{(I)}
    = \, \vec{H} \cdot \vec{J}_{t_k} + v_k^{(I)},
    \label{eq:discrete_meas}
\end{equation}
after assuming the response function of the detector, $\mathcal{R}_k(\tau)=\delta(\tau-t_k)$, to be instantaneous in each time window $t_k-\Delta<\tau\le t_k$, where by 
\begin{equation}
    v_k^{(I)} = \frac{1}{\Delta}\int_{t_k-\Delta}^{t_k} dv_\tau^{(I)},
    \label{eq:acc_det_noise}
\end{equation}
we denote discrete white noise that forms a sequence of zero-mean Gaussian variables such that $v_k^{(I)} \sim \cN(0, R^{\Delta})$ with variance $R^{\Delta}=R/\Delta$ associated with accumulated photon shot noise over each time window of size $\Delta$.  

Experimental signals are also generically affected by (i) offsets, i.e., zero-frequency contributions, and (ii) ``pink'' ($1/f$) noise, whose power increases with decreasing frequency. 
Although these can be integrated into the Gaussian noise models for $\vec{J}_t$ and $I(t)$, such complexity is mainly necessary for real-time control tasks that forbid signal preprocessing~\cite{Magrini2021}. In other cases, the structure of \eqnref{eq:spin_dynamics} ensures that the signal of interest is centred near the estimated Larmor frequency $\larmor(t)$. Hence, provided the pink-noise contribution is not large within that band, the model studied here remains applicable to real-world signals after suitable high-pass filtering~\cite{Kozbial2024}.

In this work, we simulate the operation of the spin-precession magnetometer (SPM) of 
\citet{Jimenez2018} by integrating numerically the spin dynamics \eqref{eq:spin_dynamics} and generating exemplary measurement records according to \eqnref{eq:discrete_meas}. The experimental parameters consistent with~\citeref{Jimenez2018} that we use are summarised in \tabref{tab:exp_pars}. Furthermore, by ensuring that the stated values of the atomic noise strength $Q$ and coherence time $T_2$ align with the expressions in Eqs.~\eqref{eq:Q_eff} and \eqref{eq:T_2}, respectively, we also deduce the parameters $q$, $\Gamma$, and $\alpha$ that are consistent with the experiment~\cite{Jimenez2018}---we list them below the horizontal line in \tabref{tab:exp_pars}. This allows us to investigate the impact of varying the number of atoms, $N$, in our simulations and identify the atomic density that provides the highest sensitivity. 

\newcommand{\tablepermode}{symbol}
\begin{table}[t!]
    \begin{ruledtabular}
    \begin{tabular}{lcll}
        \textrm{Name}&\textrm{Symbol}&\textrm{Value}&\textrm{Unit}\\
        \hline\noalign{\vspace{1mm}}
        Nominal Larmor freq.     &$\bar{\larmor}$ & $2\pi\times 10^4$& \SI[per-mode=symbol]{}{\radian\per \second} \\    
        Measurement strength       &$g_D$ & $0.00177$& \SI[per-mode=symbol]{}{pA}\\
        Measurement noise (var.)    &$R$&96.0& \SI[per-mode=symbol]{}{ \pico\ampere\squared\per\hertz} \\
        Number of atoms        &$N$ & $0.44\times 10^{12}$& - \\
        Atomic noise (var.)        &$Q$ & $1.26\times 10^{14}$& \SI[per-mode=symbol]{}{\hertz} \\
        Transverse coherence time      &$T_2$ & $0.87$ &\SI[per-mode=symbol]{}{ms} \\
        Sampling period  &$\Delta$ & $5$ &\SI[per-mode=symbol]{}{\micro\second}\\
        \hline\noalign{\vspace{1mm}}
        Linewidth contr.~to $\tfrac{1}{T_2}$ &$\Gamma$ & $2\pi\times 658.5$& \SI[per-mode=symbol]{}{\radian\per \second} \\
        Spin-exchange contr.~to $\tfrac{1}{T_2}$ &$\alpha$ & $2\pi\times 3.5 \times 10^{-10}$& \SI[per-mode=symbol]{}{\radian\per \second}
    \end{tabular}
    \end{ruledtabular}
    \caption{\textbf{Experimental parameters} assumed in our simulations that correspond to the spin-precession magnetometer (SPM) of~\citet{Jimenez2018}. Stemming from the relation
    \eqref{eq:T_2}, as well as~\citeref{Lucivero2016}, we deduce the parameters 
    $\Gamma$ and $\alpha$ (separated by a horizontal line) that are consistent with the experimentally observed value of $T_2$. As a result, we may study the impact of varying the atom number $N$ in our simulations.
    }
    \label{tab:exp_pars}
\end{table}

In principle, to match the initial conditions of the polarised atomic ensemble, the initial state vector $\vec{J}_0$ should be drawn from a Gaussian distribution with mean $\vec{\mu}=[0\,\tfrac{1}{2}N]^\TT$ and covariance $\E[(\vec{J}_0-\vec{\mu})(\vec{J}_0-\vec{\mu})^\TT]\ge[\frac{N}{4}\,0;0\,0]$. However, as the fluctuations of the initial state affect the subsequent dynamics and our results in a negligible manner, throughout this work we assume in our simulations the atomic spin to start from the mean, i.e., $\vec{J}_0=\vec{\mu}$. Notably, the inference methods we invoke do not possess that information, and they will be shown to be robust to the exact initial conditions.

Although we resort primarily to numerical methods to simulate the experimental conditions---this requires, however, profound numerical techniques~\cite{sarka19,kloeden92,rosler10,kuzne25} in field-tracking scenarios with time-varying $\larmor(t)$, see \appref{app:Stoch_sim}---and verify the performance of various inference tools, we also consider a simplified scenario in which the atomic noise is switched off [$Q=q=0$ in \eqnref{eq:Q_eff}] to provide analytic solutions, see \appref{app:scenario_no_atomic_noise}, in support of our numerical results.

\section{Inference methods}
\label{sec:infer}
The primary goal of Bayesian inference is to reconstruct the posterior distribution for unknown parameters of interest based on the measurement data collected. Let us denote by $\Y{k}=[\y{1},\y{2},.., \y{k}]^\TT$ the measurement record%
\footnote{
Throughout this work, we index measurement outcomes from $k=1$ with first measurement occurring after time $t_1=\Delta$.}, i.e., a sequence of $k$ consecutive measurement outcomes $\{\y{j}\in\mathbb{R}^{n_y}\}_{j=1}^{k}$, whose probability of occurrence, i.e., the \emph{likelihood function} $p(\Y{k}|\paramV)$, is determined by a vector of parameters, $\paramV\in\mathcal{D}\subset\mathbb{R}^d$, to be estimated. 
 
Then, given the \emph{prior distribution}, $p(\paramV)$, which should most accurately represent our \emph{a priori} knowledge about the parameters before taking any measurements, the \emph{posterior distribution}---the probability of the parameters taking different values given a particular measurement record---is given by the application of Bayes' rule:
\begin{equation}
    p(\paramV|\Y{k}) = \frac{p(\Y{k}|\paramV) p(\paramV)}{\int_\mathcal{D}\!d\paramV\, p(\Y{k}|\paramV)\,p(\paramV)}.
    \label{eq:posterior}
\end{equation}

Now, denoting by $\est{\paramV}(\Y{k})$ any estimator of the parameters built on the data, and choosing its average \emph{mean squared error} (MSE) as the figure of merit, i.e.,
\begin{equation}
    \Delta^2\est{\paramV}\coloneqq\tr{\errormat},
    \label{eq:MSE}
\end{equation}
where $\errormat\in\mathbb{R}^{d\times d}$ is the (average) \emph{MSE matrix} defined as
\begin{align}
    \label{eq:MSEmat}
    \errormat 
    &\coloneqq \EE{p(\paramV,\Y{k})}{(\est{\paramV}(\Y{k})-\paramV)(\est{\paramV}(\Y{k})-\paramV)^\TT} \\
    &= \!\int_\mathcal{D}\!\!\!d\paramV\,p(\paramV) \!\!\int\!\!d\Y{k}\, p(\Y{k}|\paramV)\;
        (\est{\paramV}(\Y{k})-\paramV)(\est{\paramV}(\Y{k})-\paramV)^\TT, 
    \nonumber
\end{align}
it is well known that the mean of the posterior \eref{eq:posterior}, $\est{\paramV}_\trm{opt}(\Y{k})=\int_\mathcal{D}d\paramV\,p(\paramV|\Y{k})\,\paramV$, is guaranteed to be optimal, i.e., it minimises the MSE \eref{eq:MSE}~\cite{Kay1993}.

However, as the task of computing the posterior distribution \eref{eq:posterior} is typically hard even numerically, it is common to support the analysis by computing also general lower bounds on the minimal MSE. These, when attained by a given estimator, not only turn out to be tight but also sufficient to prove the optimality of the strategy being considered. Here, we resort to the so-called \emph{Bayesian Cram\'{e}r-Rao bound} (BCRB), which holds for any estimator given a prior distribution that is smooth and possesses vanishing tails~\cite{Trees1968,Gill1995}. We use it as a benchmark to study the performance of the estimators considered.

In particular, we consider three methods of estimator construction:~the \emph{Prediction Error Method} (PEM), the \emph{Extended Kalman Filter} (EKF), and the \emph{Cubature Kalman Filter} (CKF). We argue that these methods are particularly well-suited for atomic magnetometry applications, as they all provide estimators that are constructed recursively by sequentially interpreting the measurement record~\cite{Ljung1987}. As the PEM, due to its computational complexity, is really applicable only for ``offline'' tasks with unlimited time to process the data, we present it with the intention to saturate the BCRB and prove PEM's optimality in estimating a constant magnetic field. In contrast, the Gaussian filters (EKF and CKF) by construction are designed to operate fast without large memory usage, so that despite possibly lacking optimality they are suited for ``online'' tasks, as well as tracking time-varying signals---as shown in final sections.

In what follows, after reviewing the construction of the BCRB, we present general frameworks of PEM and the nonlinear Gaussian filters (EKF and CKF), in order to then describe their application to the setting of atomic magnetometry and the SPM of interest in \secref{sub:inf_methods_spm}.

\subsection{Bayesian Cram\'{e}r-Rao Bound (BCRB)}
\label{sec:BCRB}
Given a non-pathological\footnotemark[\value{repeatedfootnote}] prior $p(\paramV)$, there exists a lower bound on the MSE matrix \eref{eq:MSEmat} that applies to any estimator (biased or unbiased), i.e., the Bayesian Cram\'{e}r-Rao Bound (BCRB)~\cite{Trees1968,Trees2007,Fritsche2014}%
\footnote{
    For any two (symmetric) matrices $\mat{A}$ and $\mat{B}$, by $\mat{A}\ge\mat{B}$ we mean that $\mat{A}-\mat{B}\ge 0$ is a positive semidefinite matrix.
}:
\begin{equation}
    \errormat
    \;\ge\;
    \errormat_\trm{BCRB} \coloneqq \BIM^{-1},
    \label{eq:BCRB}  
\end{equation}  
where
\begin{equation}
    \BIM \coloneqq
    \EE{p(\paramV,\Y{k})}{\nabla_{\paramV}\cJ(\paramV, \Y{k})(\nabla_{\paramV}\cJ(\paramV, \Y{k}))^\TT}
    \label{eq:BI}
\end{equation}
is the so-called \emph{Bayesian information} (BI)~\cite{Trees1968} expressed here in terms of the negative logarithm of the \emph{joint} distribution $p(\Y{k},\paramV)=p(\Y{k}|\paramV)\,p(\paramV)$, i.e.,
\begin{equation}
    \label{eq:log-joint}
    \cJ(\paramV, \Y{k}) \coloneqq -\ln p(\Y{k},\paramV) =-\ln p(\Y{k}|\paramV) - \ln p(\paramV),
\end{equation} 
which corresponds to the sum of the log-likelihood~\cite{Kay1993} and the contribution of the prior distribution.

Introducing the \emph{Fisher information} (FI) as~\cite{Kay1993}:
\begin{equation}
    \FIM[q_{\paramV}] \coloneqq \EE{q_{\paramV}}{(\nabla_{\paramV}\ln q_{\paramV})(\nabla_{\paramV}\ln q_{\paramV})^\TT} 
    \label{eq:FI}
\end{equation}
for any distribution $q_{\paramV}(\cdot)$ parametrised by $\paramV$, the BI \eref{eq:BI} can be conveniently rewritten as a sum of two terms:
\begin{equation}
    \BIM = \FIM[p(\paramV)]+\EE{p(\paramV)}{\FIM[p(\Y{k}|\paramV))]},
\end{equation}
which correspond to the FI evaluated w.r.t.~the prior $q_{\paramV}\equiv p(\paramV)$ and the likelihood $q_{\paramV}\equiv p(\Y{k}|\paramV)$, respectively, with the latter being also averaged over the prior.

It is important to note that the BCRB, $\errormat_\mrm{BCRB}$ in \eqnref{eq:BCRB}, does not depend on the unknown true values of the parameters $\paramV$. Still, given a particular model that determines the form of $\cJ$ in \eqnref{eq:BI}, the computation of the BCRB remains challenging, although it can often be approximated well numerically by resorting to Monte-Carlo (MC) methods. For this to be possible, however, one must be able to easily generate measurement records $\Y{k}$ for any parameter set $\paramV$ sampled from $p(\paramV)$, and importantly to numerically determine the gradient of $\cJ$ with respect to changes in $\paramV$, i.e., $\nabla_{\paramV}\cJ$, for every $\Y{k}$ generated according to a given model.

\subsection{Prediction Error Method (PEM)}
\label{sec:PEM}
The Prediction Error Methods (PEMs) are statistical algorithms designed to estimate the parameters of linear systems described by stochastic equations~\cite{Astrom1979}. Let us consider the dynamics of a system and its discrete-time measurements to be given by the following equations:
\begin{align}
    d\x{t}^{(\mrm{s})}
    &= \mat{A}_c(\paramV)\x{t}^{(\mrm{s})} dt+\mat{B}_c d\vec{w}^{(\mrm{s})}, 
    \label{eq:sys_PME}\\
    \y{k}
    &= \mat{C}\vec{x}_{k}^{(\mrm{s})}+\vec{v}_k, \label{eq:sys_output}
\end{align}
where by $\x{t}^{(\mrm{s})}\in\mathbb{R}^n$ we denote the vector representing the \emph{state} of the system (s) at time $t>0$, or by $\x{k}^{(\mrm{s})}\equiv\x{t_k}^{(\mrm{s})}$ its form at discrete time intervals $t_k=k\Delta$ with $k\in\mathbb{N}^+$. The state undergoes stochastic noise induced by the vector of mutually independent Wiener increments with unit covariance, $d\vec{w}^{(\mrm{s})}\in\mathbb{R}^{n_w}$. The matrices $\mat{A}_c(\paramV)\in\mathbb{R}^{n\times n}$ and $\mat{B}_c\in\mathbb{R}^{n\times n_w}$ are assumed to be time-invariant, with the former being also continuously differentiable w.r.t.~the parameters to be estimated, $\paramV\in \mathcal{D}\subset\mathbb{R}^d$. Moreover, the initial state is drawn from a Gaussian distribution, i.e., $\x{0}^{(\mrm{s})}\sim\cN(\vec{m}_0, \mat{P}_0)$ with known $\vec{m}_0, \mat{P}_0$. 

As in the previous section, $\y{k}\in\mathbb{R}^{n_y}$ denotes the $k$th measurement outcome, i.e., the \emph{observation}, 
which is now taken at the discrete time interval $t_k$, at which it is linearly related to the state vector via the matrix $\mat{C}\in\mathbb{R}^{n_y\times n}$. Notably, each observation is separately distorted by measurement noise, so that $\vec{v}_k\sim \cN(0, \mat{R})$ form a sequence of independent Gaussian variables. 

For measurement dynamics \eref{eq:sys_output}, it is convenient to also write the state dynamics \eref{eq:sys_PME} in discrete time as
\begin{align}
    &\vec{x}_{k}^{(\mrm{s})}=\mat{A}(\paramV)\vec{x}_{k-1}^{(\mrm{s})}+\mat{B}(\paramV)\vec{w}_{k-1}^{(\mrm{s})},
    \label{eq:pem_dt_sys}
\end{align}
where now each $\vec{w}_k^{(\mrm{s})}\in\mathbb{R}^{n_w}$ denotes an independent Gaussian variable $\vec{w}_k^{(\mrm{s})}\sim\cN(0, \1_n)$,
\begin{equation}
    \mat{A}(\paramV) \coloneqq e^{\mat{A}_{c}(\paramV)\Delta}
    \label{eq:pem_dt_A}
\end{equation}
and the matrix $\mat{B}(\paramV)$ is found as any one satisfying $\mat{B}(\paramV)\mat{B}(\paramV)^\TT=\mat{D}(\paramV)$ for the \emph{diffusion matrix} defined as
\begin{equation}
    \mat{D}(\paramV) \coloneqq \int\limits_0^{\Delta}e^{\mat{A}_{c}(\paramV)\tau}\mat{B}_c\mat{B}_c^\TT e^{\mat{A}_{c}(\paramV)^\TT\tau}d\tau.
    \label{eq:pem_dt_D}
\end{equation}

Notably, for the discrete model with \eref{eq:pem_dt_sys} and \eref{eq:sys_output} describing the state and observation dynamics, respectively, we may explicitly write the likelihood function as
\begin{equation}
    p(\Y{k}|\paramV)=\prod\limits_{j=1}^k\cN(\vec{y}_j; \mat{C}\vec{m}_j^-(\paramV), \mat{S}_j(\paramV)),
\label{eq:PEM_likelihood}
\end{equation}
with $\vec{m}_j^-$, $\mat{S}_j$ being provided by the recurrence relations that read~\cite{Sarkka,ljung1999system,söderström1989system,Bania_Baranowski_2016,Bania_Baranowski_2017}: 
\begin{subequations}
    \label{eq:pem_Kalman}
    \begin{align}
    \vec{m}_{j}^-&=\mat{A}\vec{m}_{j-1},
    \label{eq:pem_Kalman_1}\\
    \mat{P}_{j}^-&=\mat{A}\mat{P}_{j-1}\mat{A}^\TT+\mat{D},
    \label{eq:pem_Kalman_2}\\
    \mat{S}_j&=\mat{R}+\mat{C}\mat{P}_j^-\mat{C}^\TT,
    \label{eq:pem_Kalman_3}\\
    \mat{K}_j&=\mat{P}_j^-\mat{C}^\TT\mat{S}_j^{-1},
    \label{eq:pem_Kalman_4}\\
    \vec{m}_j&=\vec{m}_j^-+\mat{K}_j(\vec{y}_j-\mat{C}\vec{m}_j^-),
    \label{eq:pem_Kalman_5}\\
    \mat{P}_j&=\mat{P}_j^--\mat{K}_j\mat{S}_j\mat{K}_j^\TT,
    \label{eq:pem_Kalman_6}
    \end{align}
\end{subequations}
with the argument $\paramV$ omitted for simplicity. 

In particular, although the likelihood \eref{eq:PEM_likelihood} has a product structure, it must be evaluated in a recursive manner for a particular observation record $\Y{k}=[\y{1},\dots,\y{k}]^\TT$ with each Gaussian distribution in \eqnref{eq:PEM_likelihood} depending on the previous outcome(s) via relations \eref{eq:pem_Kalman}. In fact, for any given fixed $\paramV$ these  constitute a discrete-time Kalman filter optimal for estimating the state $\x{j}^{(\mrm{s})}$ as $\estx{j}^{(\mrm{s})}=\vec{m}_j$ with the MSE matrix \eref{eq:MSEmat} provided by \eqnref{eq:pem_Kalman_6}, i.e., $\errormat=\mat{P}_j$ after recording the $j$th measurement~\cite{Kalman1960,Kalman1961}---in the construction of $\estx{j}^{(\mrm{s})}$ \eqnsref{eq:pem_Kalman_1}{eq:pem_Kalman_2} and \eqnsref{eq:pem_Kalman_5}{eq:pem_Kalman_6} constitute prediction and correction steps, respectively, with $\mat{K}_j$ in \eqnref{eq:pem_Kalman_4} being the so-called the Kalman gain~\cite{Sarkka} (see \secref{sub:nonlinear_filters} below). 

Now, substituting the form of the likelihood \eref{eq:PEM_likelihood} into the expression \eref{eq:log-joint} and omitting all terms that are independent of $\paramV$, we obtain%
\footnote{
    For any vector $\vec{v}$ and matrix $\mat{M}$ we define $|\vec{v}|_\mat{M}^2 \coloneq \vec{v}^\TT\mat{M}\vec{v}$.
}%
\begin{align}
    \cJ(\paramV,\Y{k}) 
    &=
    \tfrac{1}{2}\sum\limits_{j=0}^k\left(|\vec{y}_j-\mat{C}\vec{m}_j^-(\paramV)|^2_{\mat{S}_j^{-1}(\paramV)}+\ln[\det\mat{S}_j(\paramV)]\right) \nonumber\\
    &\quad -  \ln p(\paramV),
    \label{eq:pem_log_like}
\end{align} 
whose minimum occurs at the same parameter values $\paramV$ (for a given measurement data $\Y{k}$) as the ones that maximise the posterior distribution $p(\theta|\Y{k})$ in \eqnref{eq:posterior}. 

As a result, with the help of PEM, we are able to construct the so-called \emph{Maximum a Posteriori} (MAP) estimator~\cite{Kay1993} by evaluating \eqnref{eq:pem_log_like} for different $\paramV$ values and solving the following minimization problem:
\begin{equation}
    \est{\paramV}_\trm{MAP}(\Y{k}) = \underset{\paramV\in\mathcal{D}}{\mathrm{argmin}}\;
    k^{-1}\cJ(\paramV, \Y{k}).
\label{eq:pem_MAP}
\end{equation}

There are many variants of the PEM in the literature~\cite{ljung1999system,söderström1989system}. In particular, in \eqnref{eq:pem_log_like} often the prior distribution $p(\paramV)$ is omitted and the matrix $\mat{S}_j$ is assumed to be constant. There also exist recursive, real-time versions of the PEM; however, they are suboptimal and based on many simplifying assumptions that may not always hold (see \citerefs{ljung1999system,söderström1989system,Bania_Baranowski_2016} for details). Notably, for large datasets ($k>>1$), the function $k^{-1}\cJ(\paramV, \Y{k})$ becomes quadratic and locally convex in the neighbourhood of its minimum. Hence, its minimum can be relatively easily found using standard optimization algorithms such as Levenberg-Maquardt, BFGS, Gauss-Newton, or Trust-Region method. A detailed description of these methods can be found in \citerefs{NoceWrig06,söderström1989system,ljung1999system}.

As an aside, let us return to BCRB \eref{eq:BCRB} and recall that in contrast to finding the minimum of \eqnref{eq:pem_log_like} with respect to $\paramV$, the computation of BI \eref{eq:BI} requires evaluation of its gradient, $\nabla_{\paramV}\cJ(\paramV, \Y{k})$, at a given point $\paramV$. Crucially, for the (linear-Gaussian) state and observation model \eqref{eq:sys_PME} and \eqref{eq:sys_output}, this can also be done in an iterative manner by formulating recursive relations similar to Eqs.~\eref{eq:pem_Kalman}, see (Thm.~A.2, p.~212) in \citeref{Sarkka}.

\subsection{Nonlinear Gaussian filters}
\label{sub:nonlinear_filters}
The Bernstein-von Mises theorem implies that the posterior \eref{eq:posterior} approaches a Gaussian distribution as the number of observations $k$ becomes asymptotically large~\cite{Vaart1998}. As a result, its maximum should then converge to the mean, and the MAP estimator \eref{eq:pem_MAP} provided by PEM should be optimal as $k\to\infty$. However, the practicality of PEM is seriously limited, as it requires recomputation of the posterior for each different $\paramV$-value considered, repeating each time the whole procedure over all past observations recorded. This makes PEM really an offline solution---although its recursive versions exist, these are cumbersome to implement and require fine-tuning of many internal parameters for a given problem~\cite{ljung1999system}. 

That is why we resort to filtering methods designed for fast estimator construction when dealing with stochastic dynamics, such as \eqnsref{eq:sys_PME}{eq:sys_output}~\cite{Sarkka}. However, for them to apply all the parameters $\paramV$ must form part of the state vector in \eqnref{eq:sys_PME} and, hence, be treated as dynamical variables. On the one hand, these can still be ensured to be time-invariant by simply setting their dynamics to be trivial, i.e., $d\paramV_t=0$. On the other, the framework opens doors to naturally deal with tasks in which the parameters $\paramV_t$ fluctuate themselves in time. In particular, it is convenient to allow them to follow analogous dynamics to the state vector in \eqnref{eq:sys_PME}, which corresponds to a (multivariate) Ornstein-Uhlenbeck (OU) process~\cite{gardiner2004handbook}: 
\begin{equation}
    d\paramV_t=\mat{M}(\paramV_t-\bar{\paramV}) dt+\mat{N} d\vec{w}_{\theta},
    \label{eq:par_dynamics}
\end{equation}
with some constant matrices $\mat{M}\in\mathbb{R}^{d\times d}$, $\mat{N}\in\mathbb{R}^{d\times d_w}$, steady-state parameter vector $\bar{\paramV}$, and independent Wiener increments of unit covariance, $d\vec{w}_\theta\in\mathbb{R}^{d_w}$, analogous to the ones in \eqnref{eq:sys_PME}.

As a result, upon defining $\x{t}\coloneq\paramV_t\oplus\x{t}^{(\mrm{s})}\in\mathbb{R}^{d'}$ and $d\vec{w}\coloneq d\vec{w}_\theta\oplus d\vec{w}^{(\mrm{s})}\in\mathbb{R}^{d'_w}$ with $d'\coloneqq d + n$ and $d'_w\coloneqq d_w + n_w$, we may now combine parameter \eref{eq:par_dynamics} and system \eref{eq:sys_PME} dynamics into a single \emph{extended} state $\x{t}$ dynamics, whereas the observation dynamics remains equivalent to the one in \eqnref{eq:sys_output}, i.e.,
\begin{align}
    d\x{t}&=\vec{f}(t,\x{t})dt+\mat{Q}d\vec{w}, \label{eq:state_dyn}\\
    \y{k}&=\mat{H}\vec{x}_{k}+\vec{v}_k, \label{eq:obs_dyn}
\end{align}
where now the vector-valued function $\vec{f}$ is \emph{nonlinear} in $\x{t}$, being determined by the matrices $\mat{M}$, $\mat{A}(\paramV)$ and $\bar{\paramV}$ in a nontrivial manner. As result, to solve the SDE \eref{eq:state_dyn}, one must resort to more advanced numerical methods~\cite{sarka19,kloeden92,rosler10,kuzne25}, which we describe in \appref{app:Stoch_sim} and use within our simulations. In contrast, the matrices $\mat{Q}\coloneq\mat{N}\oplus\mat{B}_c\in\mathbb{R}^{d'\times d'_w}$ and $\mat{H}\coloneq(\0_{n_y\times d}|\mat{C})\in\mathbb{R}^{n_y\times d'}$ just reflect the direct sum structure of the extended state.

The filtering problem, given particular extended-state \eref{eq:state_dyn} and observation \eref{eq:obs_dyn} dynamics, corresponds to determining the evolution of the posterior distribution \eref{eq:posterior} with the whole extended state $\x{t}$ being estimated, i.e., the \emph{filtering distribution}~\cite{Sarkka} defined separately in each time-interval $t\in (t_{k-1},t_{k}]$ with $k\in\mathbb{N}^+$ as
\begin{equation}
    p(t,\x{t}) \coloneq
        \begin{cases}
            p(\x{t}|\Y{k-1}),&     t_{k-1}<t<t_{k} \\
            p(\x{t_k}|\Y{k}),&     t=t_{k}
        \end{cases},
    \label{eq:filtr_distr}
\end{equation}
so that its mean can then be established to provide the optimal estimator of $\x{t}$ at time $t$, i.e., the \emph{filter}:
\begin{equation}
    \estx{t}=\int \!d\x{t}\,p(t,\x{t})\,\x{t}.
    \label{eq:nlf_x_est}
\end{equation}

The structure of \eqnref{eq:filtr_distr} implies that the filtering problem involves two steps:
\begin{enumerate}
    \item \textbf{Prediction step:} Performed in between observations by propagating the filtering distribution \eref{eq:filtr_distr} forward in time according to the Fokker-Planck equation~\cite{gardiner2004handbook} describing the state dynamics \eqref{eq:state_dyn}.
    \item \textbf{Correction step:} Performed immediately after an observation is recorded by updating the filtering distribution \eref{eq:filtr_distr} to account for the latest outcome.
\end{enumerate}
In particular, after setting the initial $p(0, \x{0})$ in \eqnref{eq:filtr_distr} to be the prior $p(\x{0})=p(\paramV_0\oplus\x{0}^{(\mrm{s})})=p(\paramV_0)p(\x{0}^{(\mrm{s})})$ describing our knowledge about the parameters and the system state at $t=0$ ($k=0$), the filtering distribution $p(t,\x{t})$ is propagated forward in time by repetitively performing prediction and correction steps over a necessary number of intervals $(t_{k-1}, t_{k}]$ with $k\in\mathbb{N}^+$, in each of which $p(t,\x{t})$ is obtained via~\cite{Sarkka}:
\begin{enumerate}
    \item \textbf{Prediction step:}\\
        For $t\in(t_{k-1},t_{k})$, the Fokker-Planck equation associated with the (It\^{o}) stochastic differential equation \eref{eq:state_dyn} is solved:
        \begin{align}
            \label{eq:nlf_fpk}
            \partial_t\, p(t,\x{t}) 
            & = -\nabla_{\vec{x}}\cdot[\vec{f}(t,\x{t})p(t,\x{t})] \\
            &\quad+\frac{1}{2}\nabla_{\vec{x}}\cdot(\nabla_{\vec{x}}\cdot[\mat{D}_c\,p(t,\x{t})]),
            \nonumber
        \end{align}
        with the (continuous-time) diffusion matrix $\mat{D}_c\coloneq\mat{Q}\mat{Q}^\TT$ and the initial condition set to $p(t_{k-1}, \x{t})$.
    \item \textbf{Correction step:}\\
        Upon recording the observation $\y{k}$ at $t=t_k$ the filtering distribution is updated as follows:
        \begin{equation}
            p(t_k, \x{t})=\frac{p(\y{k}|\x{t})p(t_k^-, \x{t})}{\int p(\y{k}|\x{t})p(t_k^-, \x{t})d\x{t}},
            \label{eq:nlf_corr_step}
        \end{equation}
        where by $p(t_k^-,\x{t})$ we denote the left-sided limit of $p(t,\x{t})$ just before recording the $k$th observation.
\end{enumerate}

Note that, according to \eqnref{eq:nlf_x_est}, the filter computes an estimate of $\vec{x}(t)$ not only at the sampling times $t_k$ but also at any time instant. This effectively increases the time resolution of measurements~\cite{Jimenez2018}. Still, due to the discrete-time nature of measurements, the density $p(t, \vec{x})$ exhibits a discontinuity at every $t_k$ with $k\in\mathbb{N}^+$. 

The above algorithm is formally optimal with the filter \eqref{eq:nlf_x_est} minimising the MSE at all times under some typical regularity conditions of $\vec{f}(t,\x{t})$ specifying the state dynamics \eqref{eq:state_dyn}. However, \eqnsref{eq:nlf_fpk}{eq:nlf_corr_step} except for very special cases---with the most prominent example of dynamics (\ref{eq:state_dyn}-\ref{eq:obs_dyn}) forming a linear-Gaussian model in which case $\estx{t}$ in \eqnref{eq:nlf_x_est} becomes the Kalman filter~\cite{Kalman1960,Kalman1961}---do not admit analytic solutions.  Furthermore, numerical methods for solving these are typically too computationally demanding for any real-time implementation.

As a result, numerous approximate methods have been developed with the most common ones relying on the \emph{Gaussian approximation}~\cite{Sarkka,Julier2004}---modelling the filtering distribution \eqref{eq:filtr_distr} to  be normally distributed at all times:
\begin{equation}
    p(t,\x{t}) \approx \cN(\x{t};\vec{m}_t,\mat{P}(t)),
    \label{eq:nlf_gauss_appr}
\end{equation}
with the mean ideally reproducing the optimal filter \eref{eq:nlf_x_est}, i.e., $\vec{m}_t\approx\estx{t}$, and the covariance matrix coinciding with the true MSE matrix \eqref{eq:MSEmat} defined now for the whole estimated state $\x{t}$, i.e., $\mat{P}(t)\approx\errormat(t)$. 

The EKF and the CKF are particular methods designed to efficiently construct $\vec{m}_t$ and $\mat{P}(t)$ in \eqnref{eq:nlf_gauss_appr} after adopting some further approximations so that the computation of the filtering distribution~\eref{eq:filtr_distr} is fast and, hence, compatible with real-time applications~\cite{Sarkka}. In this regard, one can view the (original) Kalman filter~\cite{Kalman1960,Kalman1961} as an optimal solution of a closed form---recall the likelihood \eref{eq:PEM_likelihood} and recurrence relations (\ref{eq:pem_Kalman})---applicable when the Gaussian approximation \eqref{eq:nlf_gauss_appr} is, in fact, exact thanks to the state and observation dynamics being linear and the noise being Gaussian in \eqnsref{eq:sys_PME}{eq:sys_output}. However, in the context of frequency estimation or tracking and, hence, atomic magnetometry, the ability to deal with a nonlinear $\vec{f}(t,\x{t})$ in the state dynamics \eref{eq:state_dyn} without significantly jeopardising the accuracy is essential. 

We now summarise how the EKF and CKF tackle this nonlinearity in the relevant \emph{continuous-discrete} setting of state \eref{eq:state_dyn} and observation  \eref{eq:obs_dyn} dynamics being defined in continuous and discrete time, respectively. In particular, we discuss how these algorithms decide on $\vec{m}_t$ and $\mat{P}(t)$ in \eqnref{eq:nlf_gauss_appr} in a recursive manner while scanning through the discrete measurement record. We assume the prior of the state, $p(\x{0})$, to be a Gaussian distribution of known mean and covariance:~$\vec{m}_0$ and $\mat{P}_0$, respectively; so that at $t=0$ ($k=0$) the approximation \eref{eq:nlf_gauss_appr} is exact with $p(0,\x{0})=\cN(\x{0};\vec{m}_0, \mat{P}_0)$.

\subsubsection{Continuous-discrete Extended Kalman Filter}
The continuous-discrete EKF performs the \textbf{prediction step} in each $(t_{k-1}, t_{k})$ interval by solving the following coupled differential equations for the mean and the covariance, respectively: 
\begin{subequations}
    \label{eq:nlf_ekf_pred}
    \begin{align}
        &\partial_t\vec{m}_t = \vec{f}(t,\vec{m}_t), \label{eq:nlf_ekf_pred_m}\\
        &\partial_t\mat{P}(t) = \mat{F}(t,\vec{m}_t)\mat{P}(t)+\mat{P}(t)\mat{F}(t,\vec{m}_t)^\TT+\mat{D}_c,
        \label{eq:nlf_ekf_pred_S}
    \end{align}
\end{subequations}
given the initial conditions $\vec{m}_{t_{k-1}}\equiv\vec{m}_{k-1}$ and $\mat{P}(t_{k-1})\equiv\mat{P}_{k-1}$ for a given $k\in\mathbb{N}^+$. In the above, the function $\vec{f}$ is the appearing in the state dynamics \eref{eq:state_dyn} and $\mat{F}=\nabla_{\vec{m}}\vec{f}$ is its Jacobian matrix.

Then, the \textbf{correction step} at each $t_k$ with $k\in\mathbb{N}^+$ is performed according to the following update rules:
\begin{subequations}
    \label{eq:nlf_ekf_corr}
    \begin{align}
        \mat{S}_k&=\mat{R}+\mat{H}\mat{P}_k^-\mat{H}^\TT,
        \label{eq:nlf_ekf_corr_0}\\
        \mat{K}_k&=\mat{P}_k^-\mat{H}^\TT\mat{S}_k^{-1},\\
        \vec{m}_k&=\vec{m}_k^-+\mat{K}_k(\y{k}-\mat{H}\vec{m}_k^-),\\
        \mat{P}_k&=\mat{P}_k^--\mat{K}_k\mat{S}_k\mat{K}_k^\TT.
    \end{align}
\end{subequations}
where $\vec{m}_k^-\equiv\vec{m}(t_k^-)$ and $\mat{P}_k^-\equiv\mat{P}(t_k^-)$ denote the mean and the covariance (the EKF's estimate of $\x{t}$ and its predicted error matrix), respectively, at time $t_k^-$ defined in \eqnref{eq:nlf_corr_step} that corresponds to the moment just before the $k$th observation is recorded.

\subsubsection{Continuous-discrete Cubature Kalman Filter}
The CKF differs from the EKF only by performing more accurately the \textbf{prediction step}. In particular, given the same initial conditions, $\vec{m}_{k-1}$ and $\mat{P}_{k-1}$ for each $k\in\mathbb{N}^+$, the mean and covariance are now modelled to evolve under the following equations, respectively: 
\begin{subequations}
    \label{eq:nlf_ckf_pred}
    \begin{align}
        \partial_t{\vec{m}}&=\tfrac{1}{2r}\sum\limits_{i=1}^{2r}\vec{f}(t,\vec{z}_i), 
        \label{eq:nlf_ckf_pred_m}\\
        \partial_t{\mat{P}}&=\tfrac{1}{2r}\sum\limits_{i=1}^{2r}\left(\vec{f}(t,\vec{z}_i)\vec{\xi}_i^\TT\sqrt{\mat{P}}^\TT+\sqrt{\mat{P}}\vec{\xi}_i\vec{f}(t,\vec{z}_i)^\TT\right)+\mat{D}_c, 
        \label{eq:nlf_ckf_pred_S}
    \end{align}
\end{subequations}
where $\vec{z}_i=\vec{m}_t+\sqrt{\mat{P}}\vec{\xi}_i$ with 
\begin{equation}
    \vec{\xi}_i =
    \begin{cases}
        \sqrt{r}\,\vec{e}_i, &i=1,2,\dots, r, \\
        -\sqrt{r}\,\vec{e}_{i-r}, &i=r+1,r+2,\dots,2r
    \end{cases}
    \label{eq:nlf_xi_points}
\end{equation}
where $r=d+n$ is the extended (parameter and system) state dimension with $\vec{e}_i$ denoting the $i$th basis vector in the corresponding vector space $\mathbb{R}^{r}$. In \eqnref{eq:nlf_ckf_pred_S}, the square root of $\mat{P}=\sqrt{\mat{P}}\sqrt{\mat{P}}^\TT$ is defined via the Cholesky decomposition, so that $\sqrt{\mat{P}}$ is lower triangular. For a more detailed description and the reasoning behind the decomposition \eref{eq:nlf_ckf_pred} we refer the reader to~\citerefs{SARKKA2012} and~[\citenum{Sarkka}, Chap.~6].

\subsection{Atomic magnetometry setting}
\label{sub:inf_methods_spm}
In case of the SPM described in \secref{sec:model}, the task it to estimate a single ($d=1$) parameter, in particular, the Larmor frequency $\param\equiv\larmor\in\mathbb{R}_+$. Moreover, the system state, as defined in \eqnref{eq:sys_PME}, corresponds then to nothing but the vector of relevant atomic-spin components, $\x{t}^{(\mrm{s})}\equiv\vec{J}_t$, that evolve according to the spin dynamics \eref{eq:spin_dynamics}. \eqnref{eq:spin_dynamics} then also directly specifies the two-dimensional ($n=n_w=2$) dynamical matrices in \eqnref{eq:sys_PME}:
\begin{equation}
    \mat{A}_c(\larmor)\equiv \mat{A} =
        \begin{bmatrix}
            -\frac{1}{T_2}&\larmor \\
            -\larmor&-\frac{1}{T_2}
        \end{bmatrix},
        \quad \mat{B}_c \equiv \sqrt{\mat{Q}} =
        \sqrt{Q}
            \begin{bmatrix}
                1&0 \\
                0&1
            \end{bmatrix}
    \label{eq:SPM_dynamical_mats}
\end{equation}
with $\mat{A}(t)$ in \eqnref{eq:A_Q_matrices} being time-independent when estimating a constant Larmor frequency. Recall, however, that in the above, the atomic-noise strength $Q$ is a function of the atom number, $N$, and the ensemble coherence time, $T_2$, as explained in \eqnref{eq:Q_eff}. 

For matrices \eref{eq:SPM_dynamical_mats}, the finite-step evolution of the state (spin) in between measurements over each $\Delta$ increment can be explicitly computed, so that the discrete-time equivalents of matrices \eref{eq:SPM_dynamical_mats} in \eqnref{eq:pem_dt_sys} now read:
\begin{subequations}
    \label{eq:pem_sys_matrices}
    \begin{align}
        \mat{A}(\larmor)
        &= 
        e^{-\frac{\Delta}{T_2}}
            \begin{bmatrix}
                \cos(\larmor\Delta) & \sin(\larmor\Delta)\\
                -\sin(\larmor\Delta) & \cos(\larmor\Delta)
            \end{bmatrix}, 
            \\ 
        \mat{B}
        &= \sqrt{\frac{qN}{2}\left(1-e^{-\frac{2\Delta}{T_2}}\right)}
        \begin{bmatrix}
                1&0 \\
                0&1
        \end{bmatrix}.
    \end{align}
\end{subequations}

Within the model of \secref{sec:model}, the observation dynamics \eqref{eq:sys_output} describes the photocurrent measured in discrete time steps---recall $I(t_k)$ in \eqnref{eq:discrete_meas}. Hence, each measurement outcome recorded at $t_k=k\,\Delta$ with $k\in\mathbb{N}^+$ is one-dimensional ($n_y=n_\nu=1$), so that \eqnref{eq:sys_output} takes a scalar form with $\x{t}^{(\mrm{s})}\equiv\vec{J}_t$ and the matrix $\mat{C}=[\vec{H}^\TT]$ is a row-vector, with $\vec{H}$ defined in \secref{sec:model}, i.e.,
\begin{equation}
    y_k = \mat{C}\x{k}^{(\mrm{s})}+v_k =
        \begin{bmatrix} 0 & g_D\end{bmatrix}
        \begin{bmatrix}
                J_y(t_k) \\
                J_z(t_k)
        \end{bmatrix} +v_k,
    \label{eq:SPM_obs_dyn}
\end{equation}
where the strength of measurement noise $v_k\!\equiv\! v_k^{(I)}\sim \cN(0, R^\Delta)$, recall \eqnref{eq:acc_det_noise}, is set by the photon shot noise $R^\Delta=R/\Delta$ accumulated over each time increment $\Delta$.
    
\subsubsection{PEM and BCRB}
With matrices $\mat{A}$, $\mat{B}$ and $\mat{C}$ determined, PEM can be implemented by following the steps of \secref{sec:PEM}. In particular, for any Larmor frequency $\omega$ the likelihood \eref{eq:PEM_likelihood} of a particular measurement record, $\Y{k}=\{y_1,y_2,\dots,y_k\}$, can be computed via the recurrence relations \eref{eq:pem_Kalman}. Furthermore, the logarithm of their joint distribution \eref{eq:log-joint} $p(\larmor,\Y{k})$ can be established, i.e.,~$\cJ(\larmor,\Y{k})$ in \eqnref{eq:pem_log_like}, given the prior distribution from which the Larmor frequency is drawn, $p(\larmor)=\cN(\bar{\larmor},\sigma_\larmor^2)$---assumed to be Gaussian with $\sigma_\larmor=\SI{100}{\hertz}$ for $\bar{\larmor}=2\pi \times\SI{10}{\kilo\hertz}$ in \tabref{tab:exp_pars}.   

Substituting all other parameters according to \tabref{tab:exp_pars} that we also use to simulate a given record $\Y{k}$, we construct this way numerically the MAP estimator \eref{eq:pem_MAP}. In \secref{sec:results} below, we present errors (average MSEs) of such an PEM-based approach upon averaging over $\larmor\sim\cN(\bar{\larmor},\sigma_\larmor^2)$ and sufficiently many measurement records $\Y{k}$ for each Larmor frequency value.

In parallel, we compute the BCRB \eref{eq:BCRB} that lower bounds the average MSE, $\Delta^2\est{\omega}$, in estimating a constant frequency $\omega$ and takes now a simpler form
\begin{equation}
    \Delta^2\est{\omega} 
    \quad\ge\quad 
    \frac{1}{\BI},
    \label{eq:BCRB_omega}
\end{equation}
with a scalar BI \eref{eq:BI}:
\begin{equation}
    \BI = \EE{p(\omega,\Y{k})}{(\partial_\omega\cJ(\omega, \Y{k}))^2}.
    \label{eq:BI_omega}
\end{equation}
We perform the average in \eqnref{eq:BI_omega} by resorting to MC methods and sampling over sufficiently many measurement records $\Y{k}$ for a given Larmor frequency $\omega$, being drawn randomly from the Gaussian prior distribution $p(\omega)=\cN(\bar{\larmor},\sigma_\larmor^2)$. However, dealing with only one parameter, instead of establishing a recursive ($k\to k+1$) relation to find explicitly the form of $\partial_\omega\cJ(\omega, \Y{k})$ (as described, e.g., in Ref.~[\citenum{Sarkka}, p.~212]), we use the expression for $\cJ(\omega, \Y{k})$ already determined within PEM. We then compute its derivative numerically by employing the `central difference approximation' for better convergence~\cite{NoceWrig06}. All values of the BCRB presented in \secref{sec:results} below, which contains our main results, are obtained numerically in this way.

However, to support further our findings, we establish in \appref{app:scenario_no_atomic_noise} the analytic form of the FI, BI, and the BCRB applicable in the absence of atomic noise, i.e., when $Q=0$ (and hence $\mat{B}_c=\mat{0}$) in \eqnref{eq:SPM_dynamical_mats}, which we refer to as the \emph{noiseless BCRB} for short. As it constitutes a lower bound on the true BCRB---being evaluated in a more optimistic scenario---it allows us to formally prove saturation of the MSE at long times due to finite coherence time, $T_2$, of the atomic ensemble. In particular, see \appref{app:BCRB} for details;~we prove that for any estimation strategy the average MSE \eref{eq:MSE} must always satisfy
\begin{equation}
    \Delta^2\est{\omega} \geqslant
    \left(\frac{N^2g_D^2T_2^3}{25.6R}+\frac{1}{\sigma_\larmor^2}\right)^{-1},
    \label{eq:noiseless_bcrb}
\end{equation}
where the r.h.s.~above is evaluated for $t\rightarrow\infty$ and $\Delta\rightarrow0$. 

Furthermore, as will become clear in \secref{sec:results}, because for the experimental parameters of \tabref{tab:exp_pars} the noiseless BCRB does not deviate significantly from the true BCRB also at finite probing times $t$, the former allows us to prove the $1/t^{3}$ scaling of the MSE to be exhibited at transient times. Consistently, such a cubic polynomial decay of the squared error with the sensing time (number of samples) is known to emerge at all timescales in the idealistic dissipation-free frequency estimation scenario~\cite{Rife1975}. 

\subsubsection{Nonlinear Gaussian filters}
\label{magnetometer_filters}
When constructing the Bayesian filters introduced in \secref{sub:nonlinear_filters} tailored to the atomic magnetometry setting, it becomes straightforward to generalise the scenario and allow for stochastic fluctuations of the magnetic field being sensed. In particular, the OU process postulated in \eqnref{eq:par_dynamics} now has concrete physical motivation representing the external noise~\cite{Delpy2023} imposed on the Larmor frequency $\larmor(t)$ being tracked, which evolves then according to the scalar version of \eqnref{eq:par_dynamics}:
\begin{equation}
    d\larmor(t)=-\tau^{-1}(\larmor(t)-\bar{\larmor})dt+\sqrt{d_c}dw_t^{(\larmor)},
    \label{eq:nlf_ou}
\end{equation}    
where $\tau>0$ describes the stabilisation (mean reversion) time over which the Larmor frequency converges on average to its long-term mean, $\bar{\larmor}$, whereas the strength of fluctuations is set by the diffusion coefficient, $d_c$, representing the strength of the external Gaussian noise.

Hence, the extended state $\x{t}$ introduced in \secref{sub:nonlinear_filters} that describes the parameter (Larmor frequency) and the system (atomic spin) is three-dimensional, i.e.,
\begin{equation}
    \x{t} \equiv
        \begin{bmatrix}
            x_1(t) \\
            x_2(t)  \\
            x_3(t)
        \end{bmatrix}
    =\larmor(t)\oplus\vec{J}_t= 
        \begin{bmatrix}
            \larmor(t) \\
            J_y(t)  \\
            J_z(t)
        \end{bmatrix},
        \label{eq:state_ext_spm}
\end{equation} 
and evolves according to \eqnref{eq:state_dyn} with
\begin{equation}
    \vec{f}(t,\x{t})=
            \begin{bmatrix}
                -\tau^{-1}(x_1(t)-\bar\larmor) \\
                -T_2^{-1}x_2(t)+x_1(t)x_3(t)  \\
                -T_2^{-1}x_3(t)-x_1(t)x_2(t)
            \end{bmatrix}
\end{equation}
The stochastic part of the state dynamics \eref{eq:state_dyn} is defined with the noise-increment vector $d\vec{w}=(dw_t^{(\larmor)},dw_t^{(y)},dw_t^{(z)})^\TT$ and $\mat{Q}=\mrm{diag}(\sqrt{d_C},\sqrt{Q},\sqrt{Q})$. 

Furthermore, rewriting the observation dynamics \eref{eq:SPM_obs_dyn} in terms of the extended state \eref{eq:state_ext_spm}, we obtain the desired equivalent of \eqnref{eq:obs_dyn} as
\begin{equation}
    y_k =\mat{H}\vec{x}_{k} + v_k =     
        \begin{bmatrix} 0 & 0 & g_D\end{bmatrix}
            \begin{bmatrix}
                    x_1(t_k) \\
                    x_2(t_k) \\
                    x_3(t_k)
            \end{bmatrix} +v_k,
\end{equation}
with a row-vector $\mat{H}=[0,0,g_D]$ and $v_k \sim \cN(0, R/\Delta)$.

Following \secref{sub:nonlinear_filters}, we should now effectively integrate the resulting state dynamics to obtain its discrete form $\vec{x}_k\equiv\vec{x}_{t_k}$, so that $\vec{m}_k^-$ and $\mat{P}_k^-$ in \eqnref{eq:nlf_ekf_corr} can be easily computed for each $t_k=k\Delta$ before performing the correction step of any Gaussian filter. Note that this is significantly harder than the solution presented in \eqnsref{eq:pem_dt_sys}{eq:pem_dt_D} that is valid only for linear dynamics. In particular, the necessary integration of mean and covariance for EKF \eref{eq:nlf_ekf_pred} or CKF \eref{eq:nlf_ckf_pred} may even be ill-conditioned. For the magnetometer model of interest, we find it to be time consuming enough to preclude real-time computations faster than the sampling period, $\Delta=\SI{5}{\micro\second}$ in \tabref{tab:exp_pars}.

That is why, we perform an important simplification in the filter construction, which we demonstrate to be sufficient for the frequency estimation and tracking purposes in \secref{sec:results}. Let us emphasise, however, that the true dynamics of the magnetometer is simulated without making any assumptions. In particular, when constructing the filter we assume the Larmor frequency $x_1(t)=\larmor(t)$ to change slowly enough, so that we can treat $x_1$ to be constant between sampling times in Eqs.~\eqref{eq:nlf_ekf_pred_m} and \eqref{eq:nlf_ckf_pred_m}. 

In other words, the filter is constructed as if the (extended) state dynamics was linear during the prediction step so that its discrete-time equivalent of \eqnref{eq:state_dyn} can be found following prescription of Eqs.~(\ref{eq:pem_dt_sys}-\ref{eq:pem_dt_D}) and reads
\begin{align}
    &\x{k+1}=\vec{f}(\x{k})+\mat{G}\vec{w}_k,
    \label{eq:nlf_dt_sys_0}
\end{align}
where $\vec{w}_k\sim\cN(0,\1)$ and (see also \citeref{Wang2026})
\begin{align}
    \vec{f}(\vec{x})&=
        \begin{bmatrix}
            e^{-\frac{\Delta}{\tau}}x_{1}+\bar{\omega}(1-e^{-{\frac{\Delta}{\tau}}})\\
            e^{-\frac{\Delta}{T_2}}(x_2\cos(x_{1} \Delta)+x_3\sin(x_{1}\Delta))\\
            e^{-\frac{\Delta}{T_2}}(-x_2\sin(x_{1} \Delta)+x_3\cos(x_{1}\Delta))
        \end{bmatrix},   \label{eq:nlf_dt_f}\\
    \mat{G}&=
        \begin{bmatrix}
            \sqrt{d_1}&0&0\\
            0&\sqrt{d_2}&0\\
            0&0&\sqrt{d_3}
        \end{bmatrix}, 
    \label{eq:nlf_dt_G}\\
    d_1&=\tfrac{1}{2}\tau d_c(1-e^{\frac{-2\Delta}{\tau}}),\quad d_2=d_3=\tfrac{1}{2} q N(1-e^{\frac{-2\Delta}{T_2}}).
    \label{eq:nlf_dt_di}
\end{align}

\paragraph{EKF and CKF construction.}
To construct the EKF, we set $k=0$, initialise the filter with $p(\x{0}=\cN(\vec{x},\vec{m}_0, \mat{P}_0)$ and execute the \textbf{prediction step} \eref{eq:nlf_ekf_pred} for dynamics \eref{eq:nlf_dt_sys_0}, which gives:
\begin{align}
\vec{m}_{k+1}^-&=\vec{f}(\vec{m}_{k}),\\
\mat{P}_{k+1}^-&=\mat{F}(\vec{m}_{k})\mat{P}_k\mat{F}(\vec{m}_{k})^\TT+\mat{D},
\label{eq:nlf_dt_ekf_pred}
\end{align}
where $\mat{F}=\nabla_{\vec{m}}\vec{f}$ is the Jacobian matrix for $\vec{f}$ in \eqnref{eq:nlf_dt_f}, and $\mat{D}=\mat{G}\mat{G}^\TT$ is the diffusion matrix for $\mat{G}$ in \eqnref{eq:nlf_dt_G}. 
We then perform the \textbf{correction step} following \eqnref{eq:nlf_ekf_corr}, and repeat the whole procedure recursively upon letting $k\rightarrow k+1$ until the end of the measurement record.

In case of the CKF only the \textbf{prediction step} \eref{eq:nlf_ckf_pred} is different, and it corresponds now to 
\begin{align}
\vec{m}_{k+1}^-&=\frac{1}{6}\sum\limits_{i=1}^{6}\vec{f}(\vec{z}_i),\\
\mat{P}_{k+1}^-&=\frac{1}{6}\sum\limits_{i=1}^{6} (\vec{f}(\vec{z}_i)-\vec{m}_{k+1}^-)(\vec{f}(\vec{z}_i)-\vec{m}_{k+1}^-)^\TT+\mat{D},
\label{eq:nlf_dt_ckf_pred}
\end{align}
where $\vec{z}_i=\vec{m}_{k}+\sqrt{\mat{P}_{k}}\vec{\xi}_i$ with $\vec{\xi}_i$ is given by \eqnref{eq:nlf_xi_points}. 

Both filters provide the Larmor frequency estimate at any $t_k=k\Delta$ with $k\in\mathbb{N}^+$ as $\est{\larmor}(t_k)=[\vec{m}_{k}]_1$;~however, note that the filters also provide estimates of the atomic spin components at all the discrete times.

\paragraph{Differences in performance.}
Recall that both EKF and CKF rely on the approximation of the filtering distribution \eqref{eq:filtr_distr} being a Gaussian density \eqref{eq:nlf_gauss_appr}. Moreover, EKF is optimal only if the function $\vec{f}$ in \eqnref{eq:state_dyn} is linear, whereas CKF is optimal when $\vec{f}$ is a multivariate polynomial of degree not greater than 3. There are also higher-order Gaussian approximations (e.g., Gauss-Hermite Kalman Filter~\cite{Sarkka}), but they are too complex for real-time applications. In the ideal construction, the prediction steps of the EKF \eref{eq:nlf_ekf_pred} and the CKF \eref{eq:nlf_ckf_pred} must be solved numerically using, e.g., the Runge-Kutta $4^{\mathrm{th}}$ order method or any other method appropriate for a given task. To guarantee positive definiteness of the matrix $\mat{P}$, the integration step must be sufficiently short. This can be achieved by using adaptive step size control methods. 

Here, we avoid this by linearising the prediction step as described above. Nonetheless, the discrete-time EKF requires 48 multiplications, along with a single calculation of the $\sin$ and $\cos$ functions. On the other hand, the discrete-time CKF requires 154 multiplications, triple square root calculation, and six evaluations of $\sin$ and $\cos$. We confirm that a single iteration of the EKF or CKF can be completed in less than $\SI{1}{\micro\second}$ on an ordinary personal computer. Hence, these algorithms exhibit low computational complexity and open doors to be implemented with the help of a field-programmable gate array (FPGA) to control a magnetometer in real time~\cite{Lipka2024}.

\section{Results}
\label{sec:results}
%
\subsection{Estimating constant Larmor frequency}
\label{sub:results_estimation}
\begin{figure}[t]
    \begin{center}
        \includegraphics[width=\columnwidth]{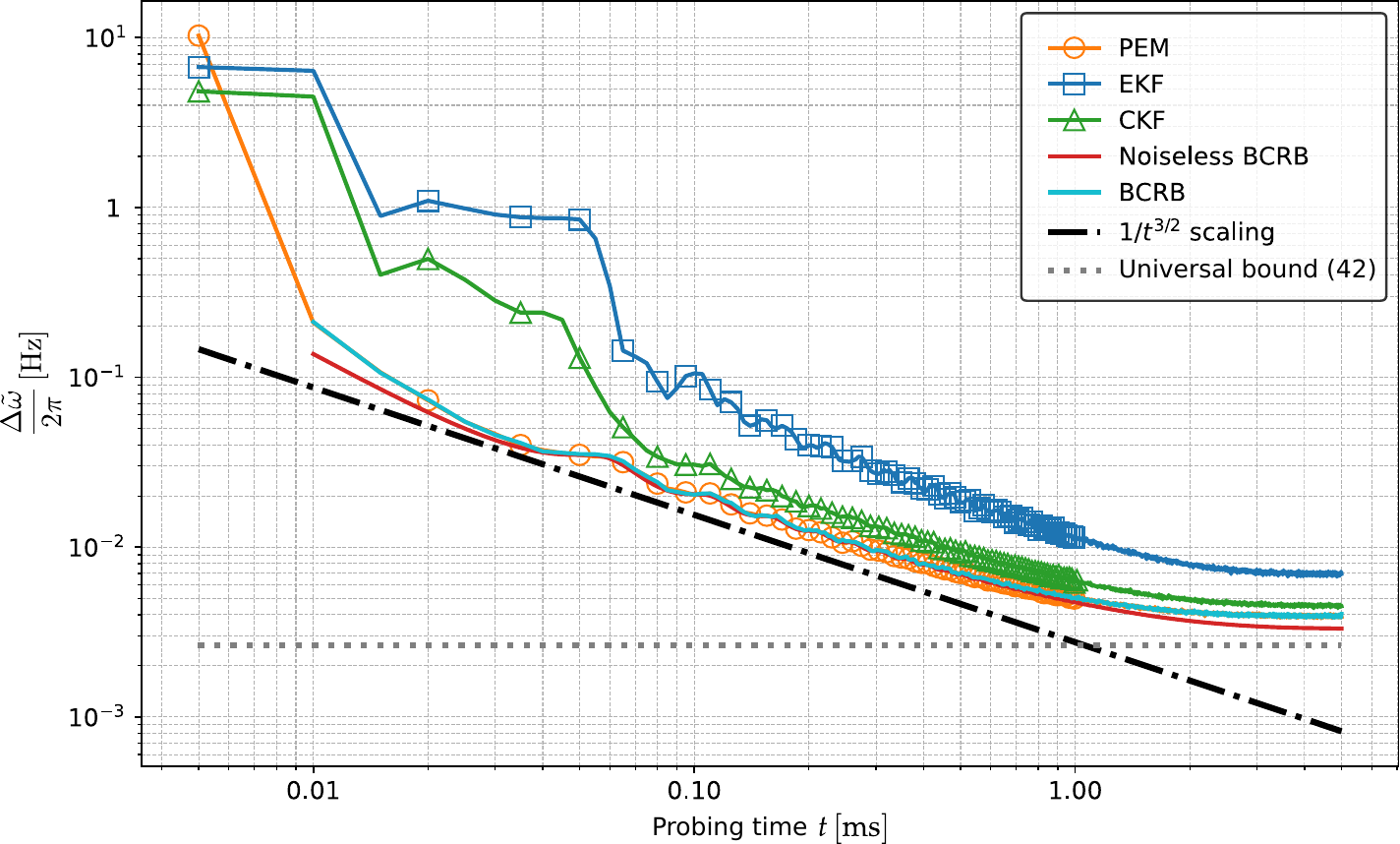}
    \end{center}
    \caption{\textbf{Estimation error $\Delta \est{\larmor}$ as a function of the probing time $t$} for estimators built using PEM (orange), CKF (green), and EKF (blue) methods. PEM is optimal, as it follows the BCRB (light blue) at all timescales---in particular, in the steady-state regime applicable beyond the coherence time ($T_2\approx\SI{1}{\milli\second}$) of the magnetometer when the minimal error is reached, whose value is constrained by the universal lower bound \eref{eq:noiseless_bcrb} (dotted grey). Both filters (EKF and CKF), however, also attain sub-$\SI{0.01}{\hertz}$ precision, reaching a spectacular relative error of $\Delta \est{\larmor}/\bar{\larmor}\lesssim\SI{e-4}{\percent}$ that is only half as precise as the BCRB. The noiseless BCRB (red) that admits an analytic form (see \appref{app:BCRB}) is also shown and is only mildly optimistic relative to the true BCRB despite assuming no atomic noise ($Q=0$). The simulation was performed with experimental parameters of \citeref{Jimenez2018} stated in \tabref{tab:exp_pars}, averaging over $10^4$ runs. For $t\gtrsim\SI{1}{\milli\second}$, we omit markers so that PEM/CKF/EKF/BCRB curves can be easily distinguished.
    }
  \label{fig:est_error_vs_time}
\end{figure}

To begin, we examine the time evolution of the estimation error for inference methods described in \secref{sec:infer}, as illustrated in \figref{fig:est_error_vs_time}. Using the parameters specified in \tabref{tab:exp_pars}, we compare the performance of PEM, EKF, and CKF to infer a constant Larmor frequency $\bar{\omega}=2\pi \times 10\SI{}{\kilo\hertz}$, i.e.,~the one used in the experiment of \citeref{Jimenez2018} and stated in \tabref{tab:exp_pars}. The plot in \figref{fig:est_error_vs_time} is obtained upon averaging simulations over $10^4$ runs, whereas all inference methods assume the same Gaussian priors:~$p(\larmor)=\cN(\bar{\larmor},\sigma_\larmor^2)$ with $\sigma_\larmor=\SI{2}{\kilo\hertz}$ for the Larmor frequency;~and $p(\vec{J})=\cN(\vec{m}_0^{(J)},\mat{P}_0^{(J)})$ with $\vec{m}_0^{(J)}=[0, N/2]^\TT$ and $\mat{P}_0^{(J)}=0.01N^2\1_2$ for the atomic spin. Note that the priors are chosen to be quite broad, allowing for large state deviations, in order to demonstrate the robustness of the inference methods employed to significant uncertainty about the initial state vector. 

The results demonstrate that PEM is optimal, in the sense that it attains the BCRB at all timescales and, hence, yields the lowest possible estimation (rms) error $\Delta \est{\larmor}$. Although suboptimal, both filters achieve sub-\SI{0.01}{\hertz} precision in the steady state at large times ($t\gtrsim T_2$), at which all errors saturate being constrained by the universal lower bound \eref{eq:noiseless_bcrb}. Still, EKF reaches a spectacular relative error of $\SI{e-4}{\percent}$ with $\Delta \est{\larmor}/\bar{\larmor}\approx 10^{-6}$ and is then approximately only half as precise as PEM. The CKF offers performance almost identical to PEM at large times, at the expense of increased computational cost.

At transient times ($t\lesssim T_2$), all estimation errors follow the $\sim 1/t^{3/2}$ dependence (dash-dotted line in \figref{fig:est_error_vs_time}), governed by the following behaviour of the MSE:
\begin{equation}
    \Delta^2\est{\larmor}\propto \frac{R}{g_D^2}\,\frac{1}{N^2\,t^{3}},
    \label{eq:t_scaling}
\end{equation}
which emerges in the absence of atomic noise ($q=0$) and can be predicted, see \appref{app:BCRB}, by examining analytically the noiseless BCRB (solid red curve in \figref{fig:est_error_vs_time}). Moreover, the asymptotic ($t\to\infty$) expression for the noiseless BCRB consistently predicts saturation of the error, which is fundamentally constrained from below by the universal bound \eref{eq:noiseless_bcrb} (dotted grey\textbf{} line in \figref{fig:est_error_vs_time}).

\begin{figure}[t]
    \begin{center}
      \includegraphics[width=\columnwidth]{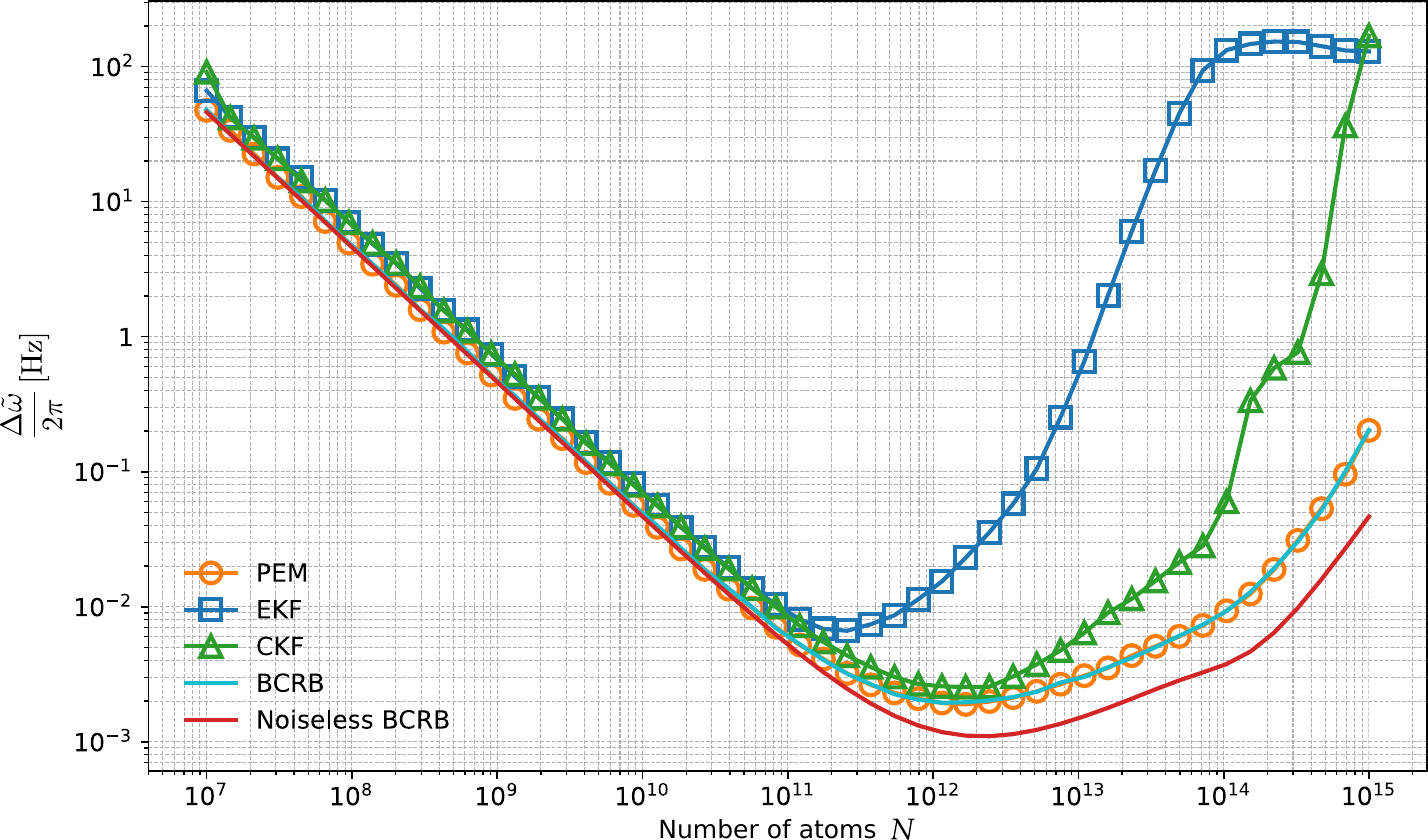}
    \end{center}
    \caption{\textbf{Estimation error $\Delta \est{\larmor}$ as a function of the atom number $N$} for estimators based on PEM (orange), CKF (green), and EKF (blue), with probing time fixed to $t=\SI{5}{\milli\second}\approx 5\,T_2$. As PEM attains again the BCRB (light blue), it forms an optimal estimator across all $N$ considered. Nonetheless, the EKF and CKF are sufficient for $N\lesssim10^{11}$ and $N\lesssim10^{12}$, respectively, at which they reach their best precision. For smaller $N$, all methods follow the noiseless BCRB \eref{eq:noiseless_bcrb} (red) and, in particular, the $1/N$-scaling that it predicts. The simulation was performed for experimental parameters of \citeref{Jimenez2018} stated in \tabref{tab:exp_pars}, averaging over $10^4$ runs.}
  \label{fig:est_error_vs_N}
\end{figure}

We further examine the behaviour of estimation errors in the steady state (for fixed $t=\SI{5}{\milli\second}\approx 5\,T_2$) as a function of the number of atoms $N$. The simulations, similar to the previous case, were averaged over $10^4$ repetitions, and the priors are the same as stated beforehand. Considering the dependence of the coherence time $T_2$ on $N$, as described in \eqnref{eq:T_2}, the results are presented in \figref{fig:est_error_vs_N}. The analysis reveals that for smaller values of $N$, the system is shot-noise limited (low atomic fluctuation strength $Q$), whereas for larger $N$, projection noise (high $Q$) becomes dominant. This is exemplified by PEM, which faithfully follows the BCRB and reaches the minimum error at $N\approx 10^{12}$, a result compatible with optimal values of atomic density observed in relevant magnetometry setups~\cite{Troullinou2023}. The divergence of error beyond this point aligns with the \emph{Signal-to-Noise Ratio} (SNR) of the spin dynamics \eref{eq:spin_dynamics}, $\trm{SNR}\coloneqq \bar{\larmor} |\vec{J}_0|/Q$, which drops rapidly around $N\approx 10^{13}$ regardless of the value of $q$ in \eqnref{eq:Q_eff} (see also \figref{fig:small_atom_noise} in \appref{app:scenario_no_atomic_noise} for $q=10^{-4}$) under the parameter conditions of \tabref{tab:exp_pars}. The CKF is able to reach the minimal error at $N\approx 10^{12}$ despite the strong atomic noise ($Q\propto N^2$ for high $N$), while the EKF begins to diverge already around $N\approx 10^{11}$. 

Since experimental setups do not necessarily operate at such high atomic densities, the trade-off between accuracy and computational complexity should be considered when choosing the estimation method.

\begin{figure}[t]
    \begin{center}
      \includegraphics[width=\columnwidth]{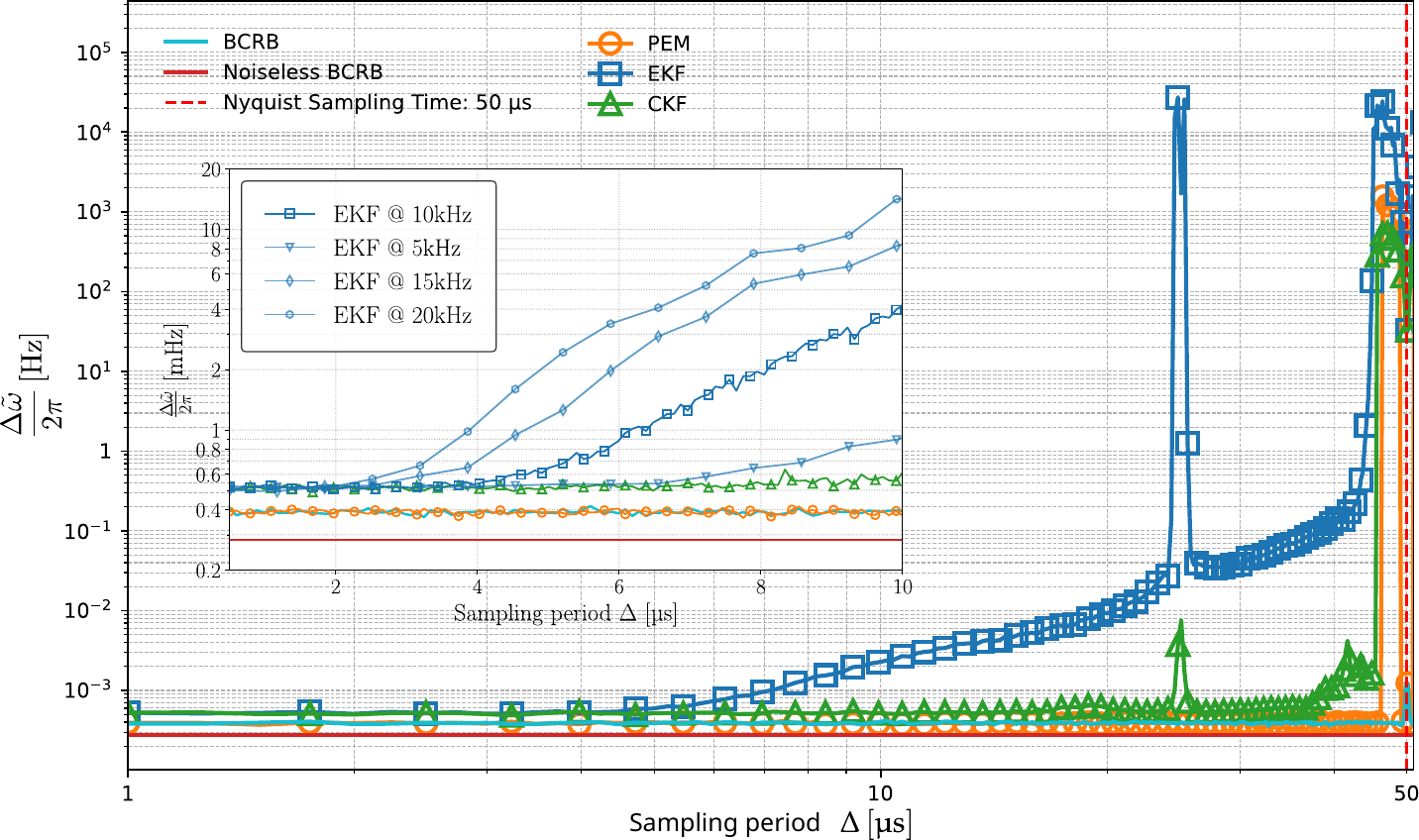}
    \end{center}
    \caption{\textbf{Estimation error $\Delta \est{\larmor}$ as a function of the sampling period $\Delta$}, achieved by the inference methods after collecting measurement data over the probing time $t=\SI{1}{\milli\second}\approx T_2$. For the Larmor frequency assumed, $\bar{\omega}= 2\pi \times \SI{10}{\kilo\hertz}$, no significant improvement for any of the methods is observed by decreasing $\Delta$ below $\approx\SI{4}{\micro\second}$. Nonetheless, the CKF is much more robust than the EKF to increasing the sampling period until becoming unstable at the Nyquist sampling period $\approx\SI{50}{\micro\second}$ (and half its value). The inset magnifies the $\Delta \in(\SI{0}{\micro\second},\SI{10}{\micro\second}]$ range and confirms that the sufficient sampling period for the EKF roughly obeys $30\,\Delta\lesssim 2\pi/\bar{\omega}$, by considering the EKF performance also when $\bar{\larmor}/2\pi=5,15,\SI{20}{[\kilo\hertz]}$. All other parameters follow \tabref{tab:exp_pars}, with the errors being averaged over $10^3$ runs.}
    \label{fig:dt_dep}
\end{figure}

Building on the previous analysis, we now investigate the impact of the measurement sampling rate on the estimation accuracy in \figref{fig:dt_dep}. Specifically, we examine how the estimation error evolves when data are collected over a fixed time interval---approximately equal to the coherence time ($t=\SI{1}{\milli\second}\approx T_2$)---while progressively reducing the sampling period $\Delta$. The results are averaged over $10^3$ repetitions using the same priors as before, with this sample size reflecting the significantly higher computational cost of performing inference at high sampling rates. 
As shown in the figure, reductions in the sampling period below $\approx\SI{4}{\micro\second}$ do not lead to a notable improvement in estimation accuracy for any of the methods. 
Conversely, as we reduce the sampling rate toward the Nyquist limit, both PEM and EKF eventually fail. While PEM continues to align with the BCRB up to the Nyquist sampling period ($\approx\SI{50}{\micro\second}$), the EKF begins to deviate much earlier, showing optimal performance only for $\Delta \lesssim \SI{5}{\micro\second}$, i.e., for about the value used in \citeref{Jimenez2018} or shorter.

To further explore this limitation, the inset in \figref{fig:dt_dep} illustrates the EKF performance across different values of the Larmor frequencies $\bar{\omega}/2\pi \in \{5, 10, 15, 20\}\, \si{\kilo\hertz}$. We observe that the "breakdown" point for the EKF shifts with the frequency, confirming that the required sampling period scales inversely with the signal $\bar{\larmor}$. Empirically, we find that a sufficient sampling period obeys the condition $30\Delta \lesssim 2\pi/\bar{\omega}$, which ensures that the filter can accurately track the signal phase within the coherence window (being dictated here by $2\pi/\bar{\omega} < T_2$).

\subsection{Tracking time-varying fields}
\label{sub:results_tracking}
Within the previous section, we have focused on the task of estimating a constant Larmor frequency in order to determine the effectiveness and accuracy of the inference methods proposed. We have benchmarked them against ultimate bounds set by the BCRB, while considering different probing times, numbers of atoms involved, as well as detection sampling rates. 

Notably, we have demonstrated that the EKF, which requires the least computational effort, is sufficient for the atomic magnetometry setting of interest, given a short enough measurement sampling period, $\Delta\lesssim\SI{5}{\micro\second}$ for the Larmor frequency of $\SI{10}{\kilo\hertz}$, and a moderate size of the atomic ensemble, $N\lesssim 10^{11}$. The EKF reaches then on average a sub-\SI{10}{\milli\hertz} resolution after a single FID of the signal, even though we have simplified the construction of the EKF even more by avoiding explicit integration over each $\Delta$ increment in \eqnref{eq:nlf_dt_sys_0}. 

That is why, in the following, we focus on the performance of solely the EKF and, in particular, study its further capabilities in tracking time-varying fields. However, as such signals may contain higher frequency components, stemming from the analysis in \figref{fig:dt_dep}, we conservatively fix the sampling period to $\Delta = \SI{1}{\micro\second}$ for the remainder of this work. This choice is further motivated by the performance of state-of-the-art data-processing units employed in current experimental setups~\cite{Lipka2024}.

\begin{figure}[t]
    \begin{center}
        \includegraphics[width=\columnwidth]{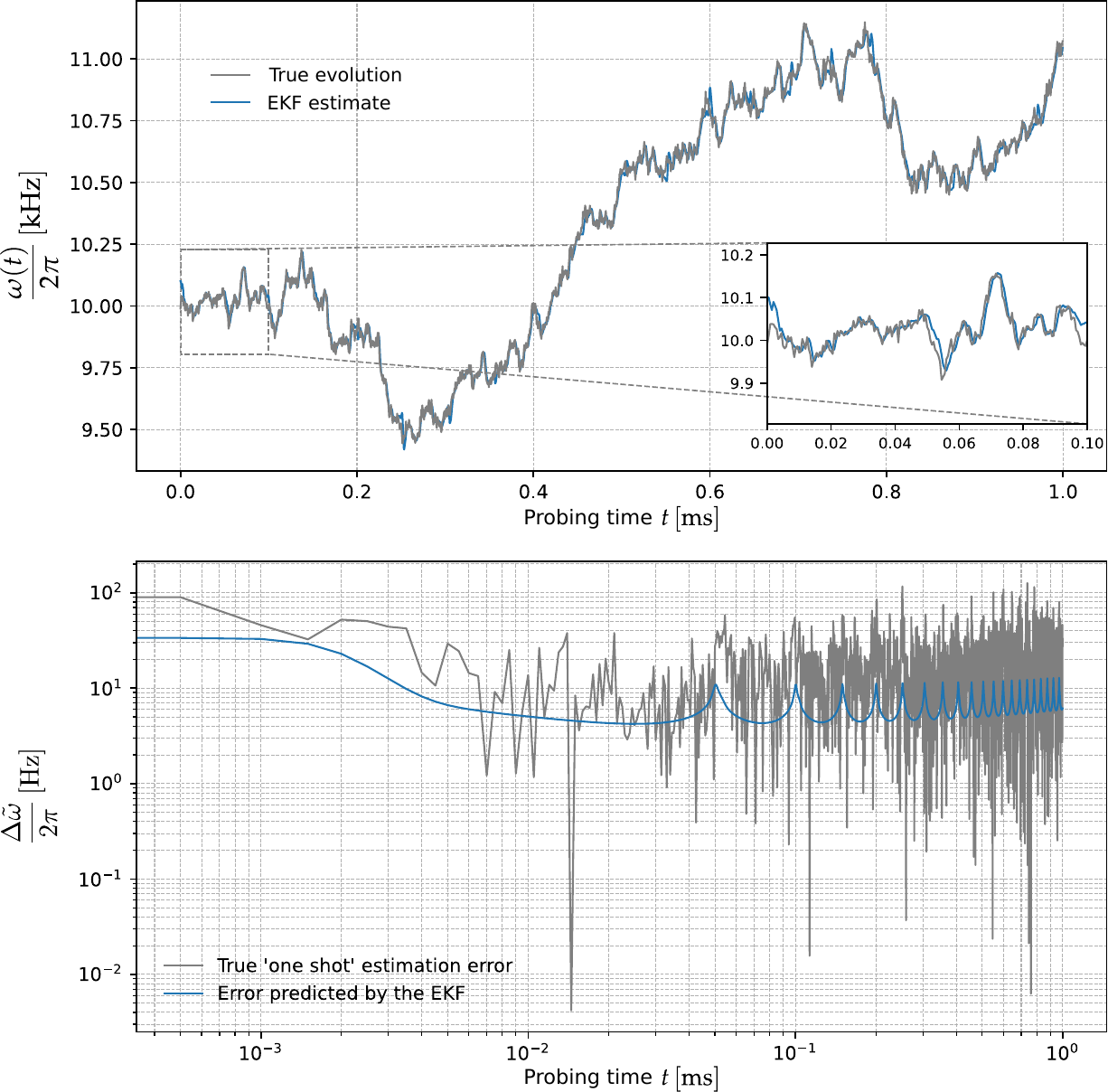}
    \end{center}
    \caption{\textbf{Tracking strong magnetic-field fluctuations using the EKF} for measurement sampling period $\Delta=\SI{1}{\micro\second}$ and other parameters of the SPM set as in \tabref{tab:exp_pars}. The Larmor frequency follows an OU process \eref{eq:nlf_ou} characterised by the mean reversion time $\tau=\SI{1}{\second}$ and the noise strength $d_c=\SI{e9}{\radian\squared\per\second\cubed}$ (top plot). For the particular experimental shot presented, the true estimation error is also shown, together with the error predicted by the EKF (bottom plot).}
  \label{fig:OU_strong}
\end{figure}

\begin{figure}[t]
    \begin{center}
        \includegraphics[width=\columnwidth]{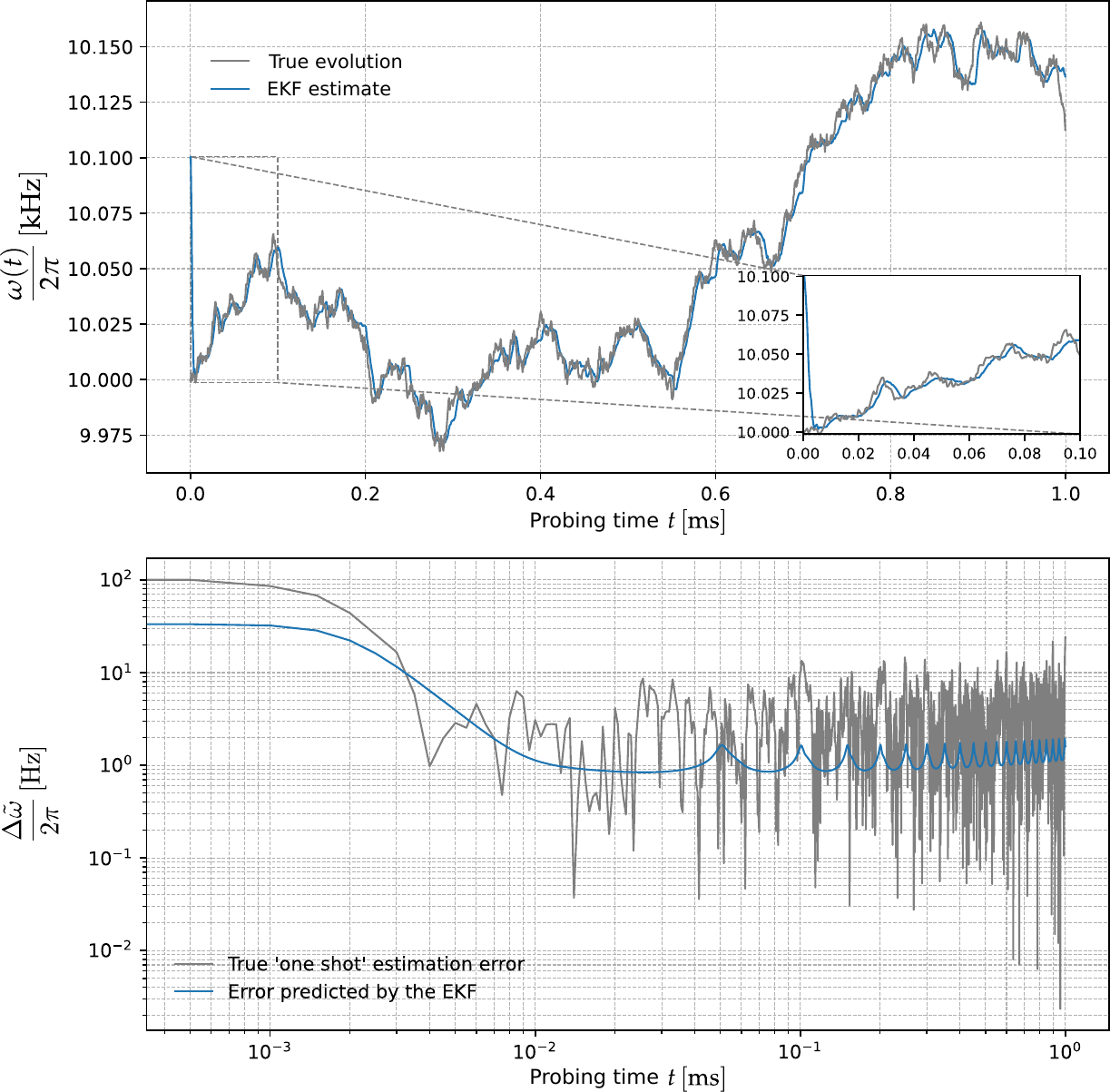}
    \end{center}
    \caption{\textbf{Tracking moderate magnetic-field fluctuations using the EKF} with measurement sampling period $\Delta=\SI{1}{\micro\second}$, and other parameters of the SPM set as in \tabref{tab:exp_pars}. The Larmor frequency follows an OU process \eref{eq:nlf_ou} characterised by mean reversion time $\tau=\SI{1}{\second}$ and noise strength $d_c=\SI{e7}{\radian\squared\per\second\cubed}$ (\emph{top plot}). For the particular experimental shot presented, the true estimation error is also shown, together with the error predicted by the EKF (\emph{bottom plot}).}
  \label{fig:OU_moderate}
\end{figure}

\subsubsection{Fluctuating field}
Firstly, we examine the estimation of the Larmor frequency for a magnetic field following the OU process introduced in \eqnref{eq:nlf_ou}, which often describes field fluctuations in atomic magnetometers operated without shielding~\cite{Delpy2023}, and is commonly used to model systems with mean-reverting evolution. In this case, the dynamical parameters of the stochastic signal $\larmor(t)$ may be easily incorporated into the EKF, recall \secref{magnetometer_filters}, which continues to perform effectively.

We consider two regimes of magnetic-field fluctuations dictated by the strengths $d_c = \SI{e9}{\radian\squared\per\second\cubed}$ and $\SI{e7}{\radian\squared\per\second\cubed}$ of the Gaussian noise in the OU process \eref{eq:nlf_ou}, which we refer to as \emph{strong} and \emph{moderate} fluctuations, respectively. Strong fluctuations correspond to an RMS variation of approximately \SI{10}{\percent} of the nominal Larmor frequency $\bar{\larmor}$ over the duration of the magnetometer's coherence time, $T_2$, whereas moderate fluctuations correspond to RMS variations of around \SI{1}{\percent}. The frequency tracking, along with the estimation errors for both cases, is shown in Figs.~\ref{fig:OU_strong}~and~\ref{fig:OU_moderate}, respectively. These results are based on single-shot experiments and demonstrate the effectiveness of the EKF in both strong and moderate fluctuation scenarios.

\begin{figure}[t]
    \begin{center}
        \includegraphics[width=\columnwidth]{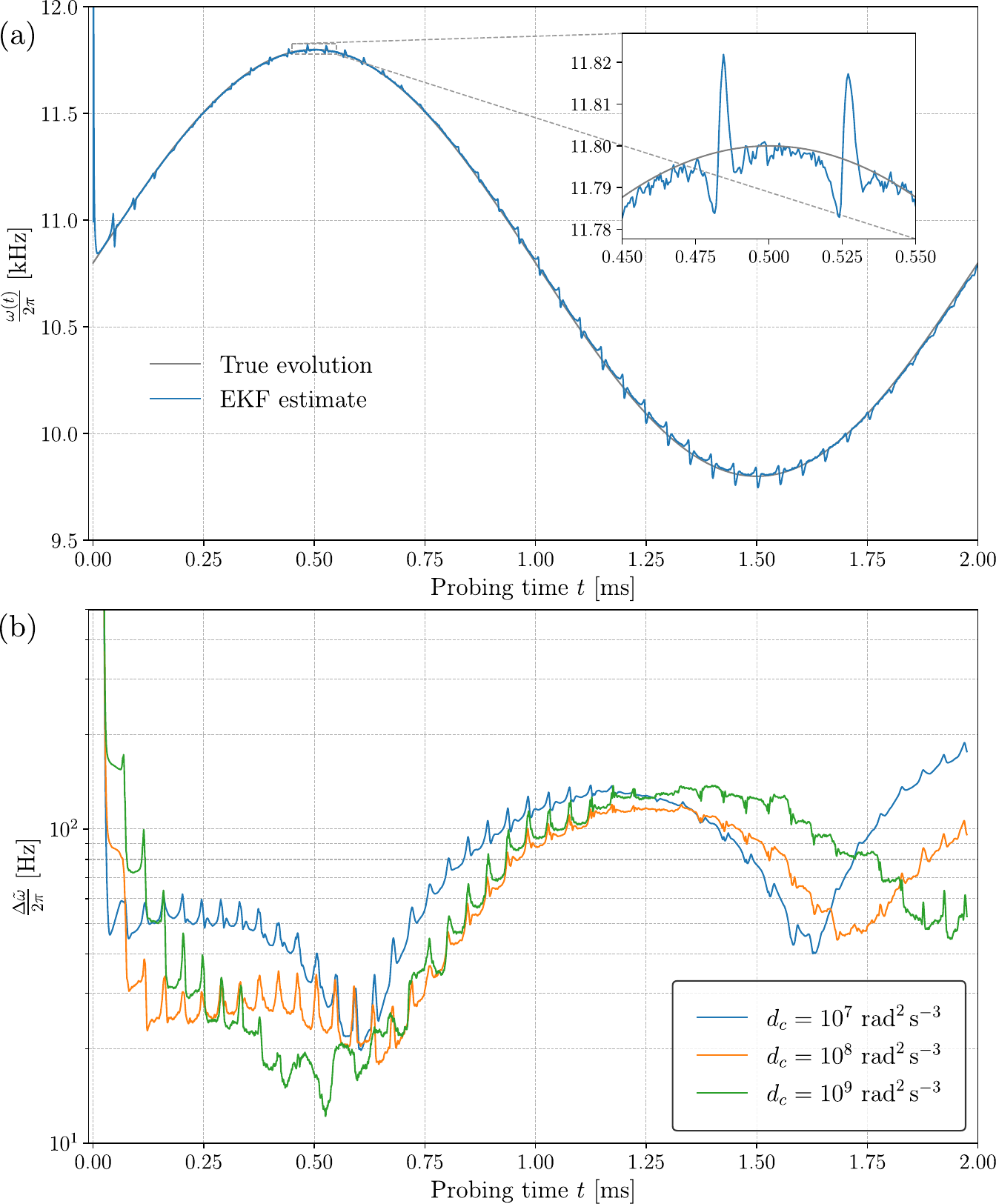}
    \end{center}
    \caption{\textbf{Tracking a sinusoidal evolution of the magnetic field using the EKF}, which expects the Larmor frequency $\larmor(t)$ to undergo Gaussian fluctuations:~a Wiener process described by \eqnref{eq:nlf_ou} with $\tau\to\infty$ and $d_c=\SI{e8}{\radian^2\second^{-3}}$. (a)~Signal reconstructed by the EKF (\emph{blue}) compared against the true signal, $\larmor(t)$ (\emph{grey}), which oscillates around $\bar{\larmor}\approx 2\pi \times \SI{10.8}{\kilo\hertz}$ with a frequency of \SI{500}{\hertz} and amplitude of \SI{1}{\kilo\hertz}. (b)~Single-shot estimation error attained by the EKF (moving average over a window size of \SI{0.1}{\milli\second}) when it expects a Wiener process of different strengths $d_c$.  It supports the choice of intermediate $d_c$ in (a), as the EKF is then ``agile'' enough to adjust quickly to the signal and its variations at short times, but also accurately follows the signal at longer timescales beyond $T_2$ without suffering excessively from the expected noise.
    }
    \label{fig:sin_waveform}
\end{figure}

\begin{figure}[t]
    \begin{center}
        \includegraphics[width=\columnwidth]{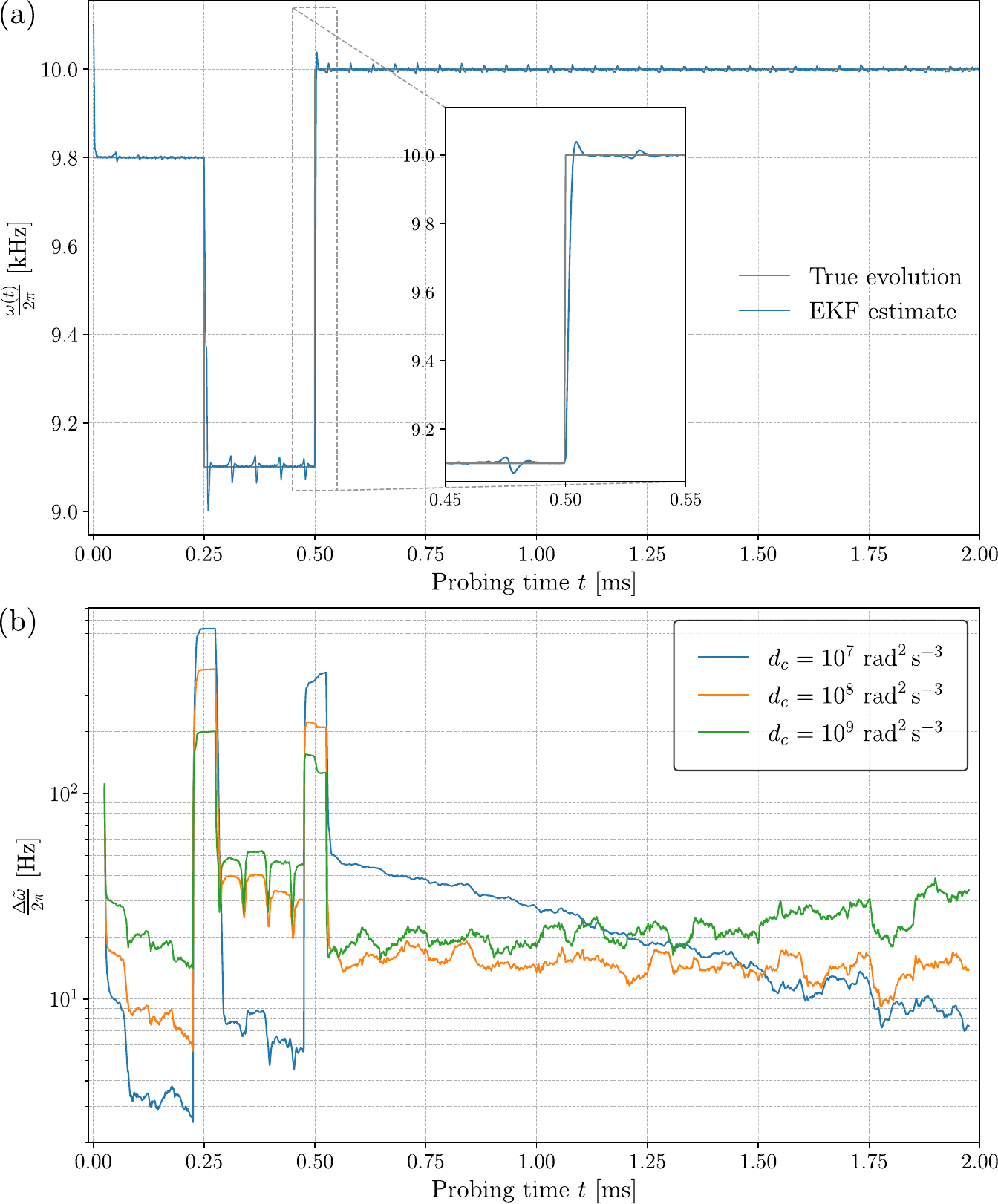}
    \end{center}
    \caption{\textbf{Tracking a step-like evolution of the magnetic field using the EKF}, which expects the Larmor frequency $\larmor(t)$ to undergo Gaussian fluctuations:~a Wiener process described by \eqnref{eq:nlf_ou} with $\tau\to\infty$ and $d_c=\SI{e8}{\radian^2\second^{-3}}$. (a)~Signal reconstructed by the EKF (\emph{blue}) compared against the true signal, $\larmor(t)$ (\emph{grey}), which instantaneously jumps from its nominal value $\bar{\larmor}\approx 2\pi \times \SI{9.4}{\kilo\hertz}$ up and down by around $\SI{0.5}{\kilo\hertz}$ over the course of $\SI{1}{\milli\second}$ ($\approx T_2$). (b)~Single-shot estimation error attained by the EKF (moving average over a window size of \SI{0.1}{\milli\second}) when it expects a Wiener process of different strengths $d_c$. As in the case of the sinusoidal waveform in \figref{fig:sin_waveform}, the choice of $d_c = \SI{e8}{\radian^2\second^{-3}}$ used in (a) is optimal. Increasing the expected noise further allows the EKF to achieve a lower error during the jump; however, this comes at the price of higher errors during periods of constant signal, which favour the lowest expected noise possible.
    }
    \label{fig:jump_waveform}
\end{figure}

\subsubsection{Oscillating and step signals}
In the OU signal scenario just described, the EKF ``knew'', i.e., was designed to optimally recover, a signal of known statistical form [\eqnref{eq:nlf_ou}] and with known parameters ($\tau$ and $d_c$).  In what follows, we employ an EKF designed to optimally recover an OU (here, Wiener) process, but use it to estimate a waveform that does \emph{not} have the OU form.  Specifically, we take $\tau\to\infty$ and set $d_c=\SI{e8}{\radian\squared\per\second\cubed}$ in \eqnref{eq:nlf_ou}, corresponding to Gaussian noise in between the strong and moderate regimes considered before.  We use it to track deterministic, rather than stochastic, time-varying changes of the Larmor frequency. This exercise gives information on the robustness of the EKF when tracking signals whose statistical form is not known \emph{a priori} and thus cannot be incorporated into the filter design. 

In particular, we examine two types of input signals:~a \emph{sinusoidal waveform} and a \emph{step function}, for which the tracking performance is presented in Figs.~\ref{fig:sin_waveform} and \ref{fig:jump_waveform}, respectively. For a slowly%
\footnote{
    Given $\larmor(t)=\bar{\larmor}+A\sin(\Omega t)$, we require $\bar{\larmor}\gg\Omega$ and $\bar{\larmor}/A\gg1$.
}
varying sinusoid---specifically at a frequency of \SI{500}{\hertz} with an amplitude of $\SI{1}{\kilo\hertz}$ (i.e., $\bar{\larmor}\pm10\%\bar{\larmor}$) in our case---the EKF performs very well, as shown in \figref{fig:sin_waveform}(a), tracking the overall shape of the waveform with only mild deviations ($\lesssim \SI{0.1}{\kilo\hertz}$) over a single FID of the SPM output signal ($t\lesssim2T_2$). Our choice of $d_c=\SI{e8}{\radian\squared\per\second\cubed}$ is supported by the single-shot error analysis presented in \figref{fig:sin_waveform}(b)---a moving average of the error is shown for a window size of $\SI{0.1}{\milli\second}$. By comparing the EKF performance after setting $d_c$ an order of magnitude less or more, we observe that decreasing $d_c$ makes the EKF too stringent to react quickly to signal variations (blue curve above yellow and green for $t\lesssim\SI{0.5}{\milli\second}$), whereas increasing $d_c$ forces the EKF to lose precision at longer timescales beyond $T_2$ (green curve above blue and yellow for $t\gtrsim\SI{1.25}{\milli\second}$), where the magnitude of noise expected by the filter dominates over the effective signal strength that decays with time.

Similarly, the EKF is capable of quickly ($\lesssim \SI{0.1}{\milli\second}$) adapting to large step-like changes of a constant magnetic field. In \figref{fig:jump_waveform}(a), we demonstrate this for jumps away from $\bar{\omega}$ on the order of \SI{0.5}{\kilo\hertz} ($\pm5\%\bar{\larmor}$), for which the EKF catches up with the frequency change over only $\approx\SI{0.02}{\milli\second}$, while experiencing quickly vanishing oscillations around the correct $\larmor$-value that arise due to nonlinearity of the dynamics. As before, we empirically verify whether the choice of $d_c=\SI{e8}{\radian\squared\per\second\cubed}$ is correct by studying the single-shot error in \figref{fig:jump_waveform}(b) (a moving average over a window size of $\SI{0.1}{\milli\second}$ is again presented). Although we observe that by increasing the expected noise $d_c$ the EKF adapts better when a jump occurs (green, yellow and blue order of curves from bottom to top during the jumps at $t=0.25,\SI{0.5}{\milli\second}$). However, this comes at the price of the highest error in between the jumps when the signal is constant, for which one should choose the lowest expected noise possible (inverted order before jumps at $t=0.25,\SI{0.5}{\milli\second}$, as well as at long timescales beyond $T_2$).

Note that, as for the OU process and Figs.~\ref{fig:OU_strong} and \ref{fig:OU_moderate}, for the time-varying deterministic signals in Figs.~\ref{fig:sin_waveform} and \ref{fig:jump_waveform}, the EKF is initialised with an initial error high enough ($\gtrsim\SI{1}{\kilo\hertz}$), so that it is clear that the necessity of the EKF to adapt to a particular signal after recording first measurement outcomes does not affect its further tracking capabilities.

In summary, the EKF proves to be a highly robust and computationally efficient tool for estimating time-varying Larmor frequencies, even when the exact nature of the field evolution is unspecified. Its ability to handle both sinusoidal variations and discrete jumps makes it a powerful approach also for scenarios where precise tracking of an unknown dynamic magnetic field is required.

\section{Conclusions} 
\label{sec:conclusions}
By simulating the operation of a spin-based quantum sensor in the free-induction decay mode, we have demonstrated that Bayesian inference methods and, in particular, the nonlinear variations of the Kalman filter can be efficiently used to estimate an external time-varying magnetic field in real time. For a constant signal, we have benchmarked our results against the ultimate bound on precision we derive, as well as studied the impact of crucial sensor parameters (size and temporal resolution) on its sensitivity. Moreover, by considering stochastic and deterministically varying waveforms we have shown our solutions to be versatile and robust. 

Although throughout our work we have assumed the sensor to consist of a fixed number of atoms $N$, this can be easily relaxed to let the atom number fluctuate from shot to shot. Given such fluctuations are well-localised around a large number of atoms---see, e.g.,~\citeref{Amoros-Binefa2025} where $N\sim\mathcal{N}(\bar N, \sigma_N^2)$ with $\bar N = 10^{13}$ and $\sigma_N=10^{11}$ is considered---this would not significantly change the results presented. On the other hand, precise determination of $N$ in experiment can be tricky, for example comparing a precise measurement of the optical attenuation against calculated per-atom optical properties~\cite{Kong2020}. For this reason, we also propose in \appref{app:natom_est} an alternative method to determine $N$, which relies on estimating the magnitude of atomic-spin fluctuations when probing the sensor in the steady state regime. See also \citerefs{MouloudakisPRA2022, MouloudakisPRA2024}.

As within our work, we have adapted a semi-classical model of an atomic magnetometer applicable for a far-detuned probing beam, a natural generalisation of our results is to consider the regime in which the measurement backaction must be accounted for~\cite{Braginsky1996,Tsang2012}. As such, a continuous quantum measurement may introduce then interatomic entanglement in the form of spin-squeezing~\cite{Kuzmich1998,Kuzmich2000} and, as our goal remains the frequency estimation rather than entanglement certification~\cite{Kong2020}, this would require a more sophisticated dynamical model in which the system state $\x{t}^{(\mrm{s})}$ in \eqnref{eq:sys_PME} encompasses also higher moments of quantum spin operators~\cite{Amoros-Binefa2024,Zhang2025}.

We emphasise that, independently of the dynamical model, our formalism is directly applicable to other types of spin-based magnetometers by just modifying the interpretation of the entries in the system state $\x{t}^{(\mrm{s})}$ in \eqnref{eq:sys_PME}---in atomic magnetometry, for instance, to ones involving atoms of higher spin or different pump-probe geometry and polarisations, e.g., alignment-based magnetometers~\cite{Ledbetter2007,Bevilacqua2014,Liu2023,Akbar2024}---as well as multiparameter sensing scenarios~\cite{Lipka2024, SierantARX2025}, e.g., vector magnetometers~\cite{Seltzer2004, Patton2014, Ingleby2018}, by simply enlarging the parameter space in the extended state $\x{t}^{(\mrm{s})}$ in \eqnref{eq:state_dyn}. 

As a result, given the efficiency of the proposed inference methods and their realization with real-time processing, e.g., using micro-controllers or FPGAs~\cite{Lee1997,Bonato2007}, we believe that our work enables real-time active feedback and control schemes in sensing scenarios involving spin-based quantum devices.

\section*{Acknowledgements}
The "Novel applications of signal processing methods in quantum sensing" project is carried out within the FIRST TEAM programme of the Foundation for Polish Science co-financed by the European Union under the European Funds for Smart Economy 2021-2027 (FENG) [P.B. and J.K.]. The research was also funded in whole or in part by the National Science Centre, Poland under grant no.~2023/50/E/ST2/00457 [K.D.]. D.M.-A., A.S. and M.W.M. acknowledge European Commission projects Field-SEER (ERC 101097313) and QUANTIFY (101135931); Horizon Europe (Q-Planet, 101291743), Chips Joint Undertaking (Chips JU) and the Spanish Ministry for Digital Transformation and Civil Service under reference CJU-010200-2026-6, as part of the Recovery, Transformation and Resilience Plan (PRTR),  Spanish Ministry of Science MCIN projects SAPONARIA (PID2021-123813NB-I00) and SALVIA (PID2024-158479NB-I00),  ``Severo Ochoa'' Center of Excellence CEX2024-001490-S [MICIU/AEI/10.13039/501100011033];  Generalitat de Catalunya through the CERCA program,  DURSI grant No. 2021 SGR 01453 and QSENSE (GOV/51/2022).  Fundaci\'{o} Privada Cellex; Fundaci\'{o} Mir-Puig. DMA acknowledges also funding from the European Union’s Horizon Europe research and innovation programme under the MSCA Grant Agreement No. 101081441.

\appendix

\section{Solving stochastic differential equations:~strong order It\^o-Taylor 1.5 scheme}
\label{app:Stoch_sim}
In order to simulate the operation of the atomic magnetometer, we must generate solutions of SDEs of the form \eref{eq:state_dyn}, which, written more concisely, reads
\begin{equation}
    d\vec{x}=\vec{f}(\vec{x})dt+\mat{Q}d\vec{w},
    \label{eq:C1}
\end{equation}
where $t\geqslant 0$, $\vec{x}(t)\in\mathbb{R}^{d'}$, $\vec{w}(t)\in\mathbb{R}^{d'_w}$  and $\mat{Q} \in\mathbb{R}^{d'\times d'_w}$. Approximating numerically the solutions to SDEs is a complex task, as standard Runge-Kutta methods are not applicable. Existing numerical algorithms fall into two main categories:~weak-order and strong-order schemes~\cite{gardiner2004handbook,kloeden92}. Weak-order schemes are primarily used to determine statistical properties of SDE solutions, but since we aim to generate realistic data for testing inference methods, strong-order schemes are required~\cite{kloeden92,rosler10,kuzne25}. The simplest strong-order approximation is the Euler-Maruyama (EM) scheme: 
\begin{equation}
    \x{k+1}=\x{k}+\Delta \vec{f}(\x{k})+\sqrt{\Delta}\mat{Q}\vec{w}_k,
    \label{eq:C2}
\end{equation}
where $\vec{w}_k\sim\cN(0, \1_{d'_w})$ and $\Delta>0$ is the time step. The EM scheme has a strong order of 1, which matches its deterministic order~\cite{kloeden92}. However, this order is too low for practical computations. 

A more accurate alternative is the It\^o-Taylor 1.5 scheme, which has a strong order of 1.5 and a deterministic order of 2~\cite{kuzne25}. This allows for larger step sizes and significantly improves computational accuracy for approximating solutions to \eqnref{eq:C1}. The scheme is given by, see e.g., Ref.~[\citenum{sarka19}, chap.~8, p.~129]:
\begin{align}
    \x{k+1}
    &= \x{k}+\Delta \vec{f}(\x{k}) + \tfrac{1}{2}\Delta^2\left(\mat{F}(\x{k})\vec{f}(\x{k})+ \vec{b}(\x{k})\right) \nonumber \\
    &\qquad +\mat{Q}\,\vec{\xi}_k+\mat{F}(\x{k})\mat{Q}\,\vec{\zeta}_k,
    \label{eq:C3}
\end{align}
where $\mat{F}\coloneqq\nabla_{\vec{x}} \vec{f} \in \mathbb{R}^{d'\times d'}$ is the Jacobian matrix of $\vec{f}$ with elements $\mat{F}_{i,j}=\frac{\partial \vec{f}_i(\vec{x})}{\partial \vec{x}_j}$. The entries (indexed by $r=1,\dots,d'$) of the vector $\vec{b}(\vec{x})\in\mathbb{R}^{d'}$ are given by
\begin{equation}
    \vec{b}_r(\vec{x})=\sum\limits_{i,j=1}^n[\mat{D}_c]_{i,j}\frac{\partial^2\vec{f}_r(\vec{x})}{\partial \vec{x}_i\partial \vec{x}_j},
    \label{eq:C4}
\end{equation}
where $\mat{D}_c=\mat{Q}\mat{Q}^\TT$ is the diffusion matrix, as defined in \eqnref{eq:nlf_fpk}. The random variables $\vec{\xi}_k,\vec{\zeta}_k\in\mathbb{R}^{d'_w}$ have a joint normal distribution with zero mean and covariance matrix: 
\begin{equation}
    \mat{S}=\begin{bmatrix}
    \Delta\; \1_{d'_w}&\tfrac{1}{2}\Delta^2\,\1_{d'_w}\\
    \tfrac{1}{2}\Delta^2\,\1_{d'_w}&\tfrac{1}{3}\Delta^3\,\1_{d'_w}
    \end{bmatrix}.
    \label{eq:C5}
\end{equation}
In practice, it is convenient to generate the variables $\vec{\xi}_k$, $\vec{\zeta}_k$ from a standard normal distribution using the following formulae:
\begin{equation}
    \vec{\xi}_k=\frac{\sqrt{\Delta}}{2}(\sqrt{3}\vec{z}_1+\vec{z}_2),
    \qquad
    \vec{\zeta}_k=\frac{\Delta^{1.5}}{\sqrt{3}}\vec{z}_1,
    \label{eq:C6}
\end{equation}
where $\vec{z}_i\sim\cN(0,\1_{n_w})$, $i=1,2$. 

As an aside, let us comment that if the matrix $\mat{Q}$ in \eqnref{eq:C1} depends on $\vec{x}$, the above scheme is not applicable.  In such cases, alternative methods of \citerefs{kloeden92,rosler10,kuzne25} should be used that are significantly more complex.

\section{Explicit solutions in the absence of atomic noise}
\label{app:scenario_no_atomic_noise}
In this appendix, the atomic magnetometry model of \secref{sec:model} is considered but in the absence of atomic noise, i.e., after setting $Q=q=0$ in \eqnref{eq:Q_eff}. In this case, the ordinary \emph{Cram\'{e}r-Rao Bound} (CRB), $1/\FI[p(\Y{k}|\param))]$, and the \emph{Bayesian Cram\'{e}r-Rao Bound} (BCRB), $1/\BI[p(\Y{k},\param))]$ [scalar version of \eqnref{eq:BCRB}], are derived, providing fundamental bounds on the precision of frequency estimation. Although the CRB applies to \emph{frequentist} estimation problems (sufficiently many independent repetitions)~\cite{Kay1993} instead of \emph{Bayesian} ones (one shot) that are of relevance to our work and atomic magnetometry, its analysis is still insightful being independent of the prior distribution $p(\theta)$, whose choice may strongly affect the BI \eref{eq:BI} and, hence, the BCRB \eref{eq:BCRB}. Moreover, once the CRB is determined, which applies only to (locally) \emph{unbiased} estimators~\cite{Kay1993}, the BCRB can be easily computed for a specific $p(\theta)$, which applies to \emph{all} estimators given $\lim_{\theta\to\pm\infty}p(\theta)=0$~\cite{Trees1968,Trees2007,Fritsche2014}.

The analytic form of the CRB is analysed in the limit of a high sampling rate ($\Delta\rightarrow0$) and large observation times ($t\rightarrow\infty$). Following this, the scenario of long coherence time $T_2$ is specifically examined, where the signal exhibits slower decay and more spin oscillations can be observed over the given interval. The expressions for the CRB are explicitly compared against the performance of inference methods introduced in the main text (PEM, EKF, and CKF) when applied to measurement data simulated for the relevant idealistic magnetometry characteristics.

Finally, we also compute the corresponding BCRB for a Gaussian prior distribution $p(\theta)$, but also upper bound it irrespective of the prior. We then numerically examine both CRB and BCRB as functions of time, in order to explicitly verify their analytically predicted scalings of $\propto 1/t^{5}$ and $\propto 1/t^{3}$ that emerge at short and transient times, respectively, before the impact of finite (transverse) coherence time, $T_2$, starts to play the role.

\subsection{Fisher information and the CRB}
As explained in \secref{sub:inf_methods_spm}, the SPM dynamics \eref{eq:spin_dynamics} is just an instance of the SDE \eref{eq:sys_PME} with dimensions $n=n_w=2$ and matrices $\mat{A}_c$, $\mat{B}_c$ specified in \eqnref{eq:SPM_dynamical_mats}. Now, in the absence of atomic noise ($\mat{B}_c=0$), \eqnref{eq:sys_PME} becomes just an ordinary differential equation with solution:
\begin{equation}
    x_1(t)=\tfrac{N}{2}e^{-\frac{t}{T_2}}\sin(\theta t),     
    \qquad
    x_2(t)=\tfrac{N}{2}e^{-\frac{t}{T_2}}\cos(\theta t),
    \label{eq:state_dyn_no_at_noise}
\end{equation}
given the initial conditions $\vec{x}(0)=\begin{bmatrix}0&N/2\end{bmatrix}^\TT$ dictated by the initial polarisation of the atomic ensemble ($\vec{J}_0=\vec{\mu}$). 

As a result, each observation in \eqnref{eq:SPM_obs_dyn} with $j\in\mathbb{N}^+$ and $t_j=j\Delta$ reads
\begin{equation}
    y_j=\tfrac{N}{2}e^{-\frac{t_j}{T_2}}\cos(\theta t_j)+v_j,
    \label{eq:obs_dyn_no_at_noise}
\end{equation}
where $v_j\sim\cN(0,\sigma_v^2)$ are independent random variables modelling the measurement noise, whose strength $\sigma_v^2=R/(g_D^2\Delta)$ we renormalise here for convenience. 

Defining $\varphi(\theta, t)\coloneqq\tfrac{N}{2}e^{-\frac{t}{T_2}}\cos(\theta t)$ as the deterministic part of \eqnref{eq:obs_dyn_no_at_noise}, we explicitly write the corresponding (Gaussian) likelihood \eref{eq:PEM_likelihood} of obtaining a particular measurement record, $\Y{k}=\{y_1,y_2,\dots,y_k\}$, as
\begin{equation}
    p(\Y{k}|\theta)=\prod\limits_{j=1}^{k}\cN(y_j; \varphi(\theta, t_j), \sigma_v^2),
    \label{eq:A3}
\end{equation}
and its negative log-likelihood function:
\begin{equation}
    \cL(\theta, \Y{k})=\frac{k}{\sqrt{2\pi}\sigma_v}+\frac{1}{2\sigma_v^2}\sum\limits_{j=1}^{k}(\varphi(\theta, t_j)-y_j)^2.
    \label{eq:A5}
\end{equation}
Consequently, defining the scalar version the FI \eref{eq:FI} as
\begin{align}
    \FI[p(\Y{k}|\theta)] 
    &\coloneqq \EE{p(\Y{k}|\theta)}{(\partial_{\param}\ln p(\Y{k}|\theta))^2} \nonumber \\
    &= \EE{p(\Y{k}|\theta)}{(\partial_{\param}\cL(\theta, \Y{k}))^2}
    \label{eq:scalarFI}
\end{align}
we obtain its explicit form for our problem:
\begin{align}
    \FI[p(\Y{k}|\theta)]    
    &= \frac{N^2}{4\sigma_v^2}\sum\limits_{j=1}^{k}e^{-\frac{2t_j}{T_2}}t_j^2\sin^2(\theta t_j) \nonumber\\
    &= \frac{N^2g_D^2}{4R}\Delta\sum\limits_{j=1}^{k}e^{-\frac{2t_j}{T_2}}t_j^2\sin^2(\theta t_j).
    \label{eq:A10}
\end{align}

Now, considering the continuous-time limit ($\Delta\to0$) of high sampling rates by fixing $t=k\Delta t$ and letting simultaneously $k\rightarrow\infty$, we can replace the sum above by an integral. Denoting by $\Y{t}$ the  continuous measurement record thus obtained, its FI \eref{eq:FI} reads
\begin{equation}
    \FI[p(\Y{t}|\larmor)]=\frac{N^2g_D^2}{4R}\int\limits_{0}^{t}e^{-\frac{2t}{T_2}}t^2\sin^2(\larmor t)dt,
    \label{eq:FI_no-noise}
\end{equation}
where we have also substituted for the (constant) Larmor frequency $\theta\equiv\larmor$ being the estimated parameter in the atomic magnetometry context.

Although we derive an exact formula for the integral in \eqnref{eq:FI_no-noise} that may be interpreted as the continuous-time limit of a damped sinusoidal process~\cite{Yao1995}, it is too cumbersome to be included here. Instead, we present its expansion for very short observation times ($t\approx0$):
\begin{equation}
    \FI[p(\Y{t}|\larmor)]=\frac{g_D^2N^2\larmor^2}{20R}t^5+O(t^5),
\label{eq:FI_no-noise_short_time}
\end{equation}
as well as its asymptotic value evaluated for infinitely large observation time ($t\rightarrow\infty$):
\begin{equation}
    \FI[p(\Y{t}|\larmor)]
    \underset{t\to\infty}{=}
    \frac{N^2g_D^2T_2^3}{32R}\frac{(\larmor T_2)^2((\larmor T_2)^4+3(\larmor T_2)^2+6)}{(1+(\larmor T_2)^2)^3}.
    \label{eq:FI_no-noise_asympt}
\end{equation}
The expression \eref{eq:FI_no-noise_short_time}, which exhibits the $t^5$ scaling with time, applies in the effective regime of no decoherence, being independent of the transverse coherence time $T_2$. In contrast, the emergence of a constant asymptotic value \eref{eq:FI_no-noise_asympt} is a direct consequence of the coherence time being finite, with \eqnref{eq:FI_no-noise_asympt} consistently vanishing as $T_2\to0$.

Finally, let us comment that the above behaviour of the FI that we derive is \emph{not} in conflict with the well-established results of frequency estimation~\cite{Kay1993}, which state that for sinusoidal signals with Gaussian noise the FI scales cubically with the number of samples, i.e., $k^3=(t/\Delta)^3$~\cite{Rife1975}. In particular, note that in the limit of no atomic decoherence, $T_2\to\infty$, reads
\begin{align}
    \label{eq:FI_no-noise_no-dec}    
    \FI[p(\Y{t}|\larmor)]
    & =\frac{g_D^2N^2}{4R}\int\limits_{0}^{t}\tau^2\sin^2(\larmor \tau)d\tau \\
    & = \frac{g_D^2N^2}{24R} t^3\left(1+\frac{3a(t)}{4(\larmor t)^3}\sin(2\larmor t+\phi(t))\right),
    \nonumber
\end{align}
where 
\begin{equation}
    a(t) \coloneqq \sqrt{(1-2(\larmor t)^2)^2+4(\larmor t)^2}
\end{equation}
and $\phi(t)$ is determined as the solution to
\begin{equation}
    \cos\phi(t)=\frac{1-2(\larmor t)^2}{a(t)}
    \quad\text{and}\quad
    \sin\phi (t)=\frac{-2\larmor t}{a(t)}.
\end{equation}
Notably, it is not hard to prove that the FI \eref{eq:FI_no-noise_no-dec} consistently reproduces the expansion \eref{eq:FI_no-noise_short_time} at short observation times, whereas for large $t$ the term in curly brackets in \eqnref{eq:FI_no-noise_no-dec} can be ignored, so that
\begin{equation}
    \FI[p(\Y{t}|\larmor)] \underset{t\gg1}{\approx} \frac{g_D^2N^2}{24R} t^3,
\label{eq:FI_no-noise_no-dec_larget}
\end{equation}
and the cubic scaling is consistently retrieved~\cite{Rife1975}. 

Although \eqnref{eq:FI_no-noise_no-dec_larget} applies in the absence of decoherence ($T_2=0$), it allows us to predict together with the whole above analysis that the CRB~\cite{Kay1993},
\begin{equation}
    \Sigma_\trm{CRB} \coloneqq \frac{1}{\FI[p(\Y{t}|\larmor)]},
    \label{eq:CRB}
\end{equation}
should scale with the operation time as $t^{-5}$ and $t^{-3}$ at short and transient times, respectively, before saturating at the asymptotic value \eref{eq:FI_no-noise_asympt} imposed by the decoherence $T_2>0$. We show this explicitly in what follows, however, let us note that the presence of atomic noise, i.e., $q>0$ in \eqnref{eq:Q_eff}, can only worsen (i.e., increase) the CRB \eref{eq:CRB} even further. Hence, \eqnsref{eq:FI_no-noise_asympt}{eq:FI_no-noise_no-dec_larget} apply and support the results of the main text and \secref{sub:results_estimation}, in which we explicitly observe for the SPM of \citeref{Jimenez2018} (see \figref{fig:est_error_vs_time}) the transition of the MSE from the (noiseless) $t^3$-scaling to the decoherence-induced asymptotic costant.

\begin{figure}[t]
    \begin{center}
      \includegraphics[width=\columnwidth]{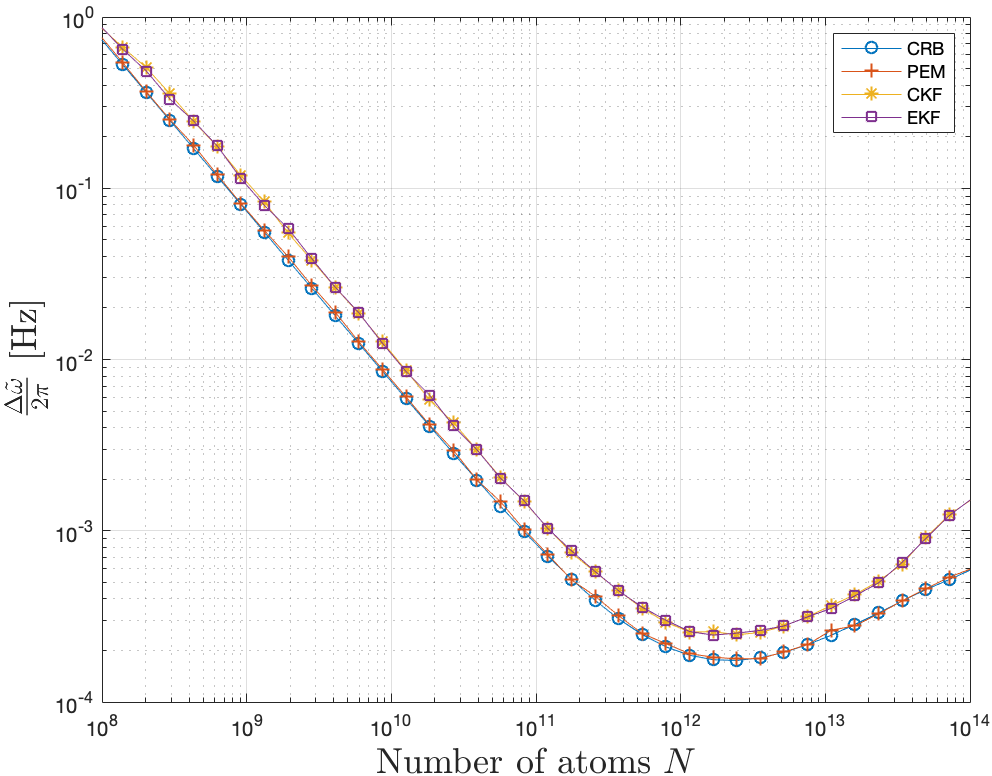}
    \end{center}
    \caption{\textbf{Estimation error as a function of the number of atoms $N$ for small atomic noise}:~$q=10^{-4}$ in \eqnref{eq:Q_eff}. All other parameters are set as in \tabref{tab:exp_pars}, and probing time $t=\SI{5}{\milli\second}\approx 5\,T_2$ is considered. Estimators are computed employing Bayesian methods:~PEM (red crosses), CKF (yellow stars) and EKF (violet squares); assuming Gaussian priors for the Larmor frequency, $\larmor\sim\cN(\bar{\larmor},\sigma_\larmor)$ with $\bar{\larmor}= 2\pi \times \SI{10}{\kilo\hertz}$ and $\sigma_\larmor = \SI{10}{\hertz}$, and the initial atomic spin, $\vec{J}_0 \sim \cN(\vec{\mu},\mat{P}_0)$ with $\vec{\mu}=[0, N/2]$ and $\mat{P}_0=qN^2\,\1_2$. These are compared against the CRB \eref{eq:CRB} evaluated with no atomic noise ($Q=0$) via \eqnref{eq:FI_no-noise_asympt}. Note that in case of the atomic noise being small, PEM saturates the above CRB for all atom numbers, whereas the errors attained by EKF and CKF are almost identical.}
    \label{fig:small_atom_noise}
\end{figure}

\subsection{Simulation and inference performance}
We repeat simulations for the magnetometer of interest, but significantly lower the atomic noise, i.e., the value of $q$ stated in \tabref{tab:exp_pars}, for the expression derived above (for $q=Q=0$) to be applicable. In \figref{fig:small_atom_noise}, we set $q=10^{-4}$ and present the estimation error achieved by each of the Bayesian inference methods (assuming a narrow Gaussian prior) introduced in the main text (PEM, EKF and CKF) as a function of the number of atoms $N$. 
Notably, we observe that for all $N$ PEM-based estimator saturates the CRB evaluated assuming $Q=0$ by computing \eqnref{eq:FI_no-noise_asympt}. Moreover, the errors of EKF and CKF are nearly identical and only slightly larger than the PEM error.

\begin{figure}[t]
    \begin{center}
      \includegraphics[width=\columnwidth]{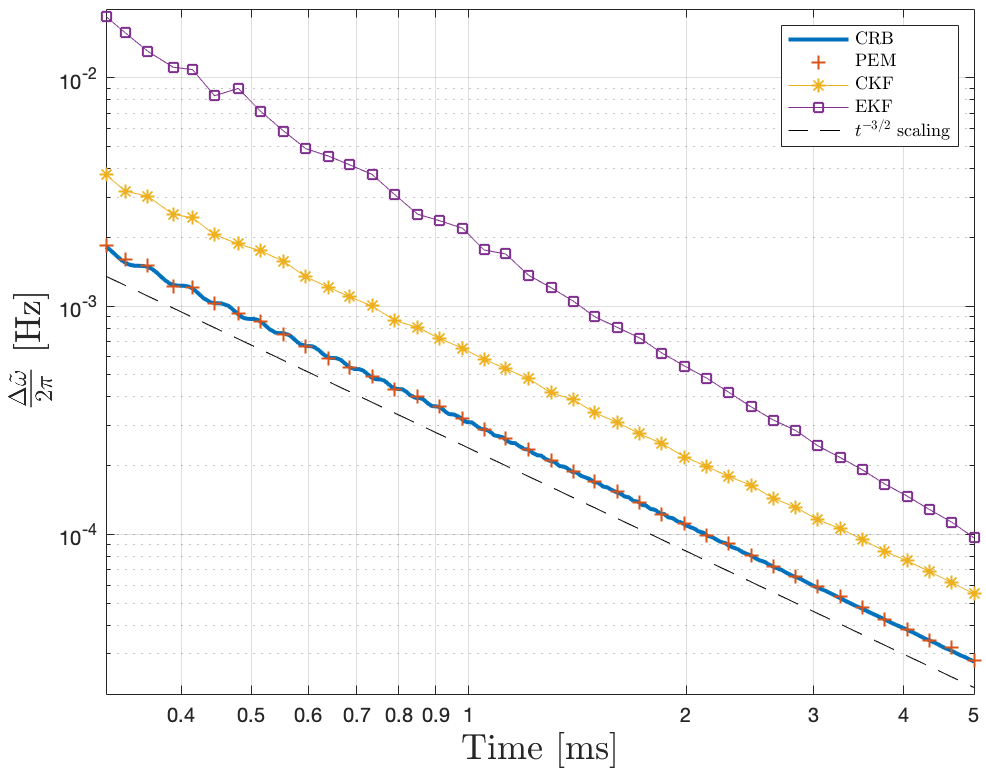}
    \end{center}
    \caption{\textbf{Estimation errors attained over time, compared against the CRB for weak atomic decoherence} ($T_2=\SI{1}{s}$). All other parameters, apart from $Q=0$, are set as in \tabref{tab:exp_pars}. PEM (red crosses), CKF (yellow stars) and EKF (violet squares) methods are considered, whereas the CRB (solid blue) is computed assuming $T_2\to\infty$ with help of \eqref{eq:FI_no-noise_no-dec}. In agreement with \eqnref{eq:FI_no-noise_no-dec_larget} (dashed black), it is clear that the average MSE scales as $t^{-3}$ with the detection time. All three Bayesian estimators are constructed assuming Gaussian priors for the Larmor frequency, $\larmor\sim\cN(\bar{\larmor},\sigma_\larmor)$ with $\bar{\larmor}= 2\pi \times \SI{10}{\kilo\hertz}$ and $\sigma_\larmor = \SI{2}{\kilo\hertz}$,  and the initial atomic spin, $\vec{J}_0 \sim \cN(\vec{\mu},\mat{P}_0)$ with $\vec{\mu}=[0, N/2]$ and $\mat{P}_0=(qN^2/20)\1_2$.
    }
    \label{fig:large_T2}
\end{figure}

Furthermore, in \figref{fig:large_T2} we consider the magnetometer operating in the regime of low atomic decoherence, i.e., large transverse coherence time $T_2$. Notably, in this regime, the number of spin oscillations observed within the interval $[0, T_2]$ is large. As a result, the signal carries substantially more information (compare against \figref{fig:est_error_vs_time} of the main text) about the Larmor frequency, leading to a rapid decrease in the squared estimation error of PEM, CKF and EKF as $t^{-3}$, predicted by \eqnref{eq:FI_no-noise_no-dec_larget}.

\subsection{BCRB and its $t$ dependence}
\label{app:BCRB}
According to the general theory presented in \secref{sec:infer} (c.f.~\citeref{Trees1968}), the BCRB \eref{eq:BCRB} ensures for a scalar parameter $\larmor$ that the average MSE for any (also biased) estimator fulfils the inequality $\Delta^2\est{\larmor}\ge \Sigma_\trm{BCRB}$, where
\begin{equation}
    \Sigma_\trm{BCRB} \coloneq \frac{1}{\BI} = \frac{1}{\FI[p(\larmor)]+\EE{p(\larmor)}{\FI[p(\Y{k}|\larmor))]}}.
    \label{eq:A6}
\end{equation}

Now, assuming a Gaussian prior distribution of the true frequency value, i.e., $p(\larmor)=\cN(\larmor;\bar{\larmor},\sigma_\larmor)$, we get
\begin{equation}
    \FI[p(\larmor)]=\sigma_\larmor^{-2},
    \label{eq:A9}
\end{equation}
whereas the part of BI associated with the observation, whose FI is given by \eqnref{eq:FI_no-noise}, reads
\begin{equation}
    \EE{p(\larmor)}{\FI[p(\Y{k}|\larmor)]} = \left\langle \frac{N^2g_D^2}{4R}\int\limits_{0}^{t}e^{-\frac{2t}{T_2}}t^2\sin^2(\larmor t)dt \right\rangle_{p(\larmor)}.
    \label{eq:J_D_exact}
\end{equation}

Moreover, under the assumption of a Gaussian prior, the above integral admits an exact analytical evaluation; however, the resulting expression is extremely intricate. Consequently, we proceed by deriving a useful upper bound for \eqref{eq:J_D_exact}. We consider the limit $t\rightarrow\infty$, in which we may substitute \eqnref{eq:FI_no-noise_asympt} and obtain
\begin{align}
    & \EE{p(\larmor)}{\FI[p(\Y{k}|\larmor)]} 
    = \frac{N^2g_D^2T_2^3}{32R} \times  \label{eq:JD}\\
    & \qquad \times \int p(\larmor)\frac{(\larmor T_2)^2((\larmor T_2)^4+3(\larmor T_2)^2+6)}{(1+(\larmor T_2)^2)^3}d\larmor.
    \nonumber
\end{align}
Since the above integral is not greater than $\tfrac{5}{4}$ we have
\begin{equation}
    \EE{p(\larmor)}{\FI[p(\Y{k}|\larmor)]} \leqslant \frac{N^2g_D^2T_2^3}{25.6R}.
    \label{eq:JD_upper}
\end{equation}

Consequently, combining Eqs.~\eqref{eq:A6}, \eqref{eq:A9} and \eqref{eq:JD_upper} we obtain the inequality \eqref{eq:noiseless_bcrb} stated in the main text. However, let us note that for a realistic scenario, i.e., for the parameters given in \tabref{tab:exp_pars} and $N\geqslant 10^{8}$ the contribution of the prior to the BCRB can be safely omitted by setting $\sigma_\larmor\to\infty$. For instance, given $\sigma_\larmor\geqslant 2\pi\times \SI{1}{\kilo\hertz}$, the ratio $\EE{p(\larmor)}{\FI[p(\Y{k}|\larmor)]}/\FI[p(\larmor)]$ is more than $10^{4}$.

\begin{figure}[t]
    \begin{center}
      \includegraphics[width=\columnwidth]{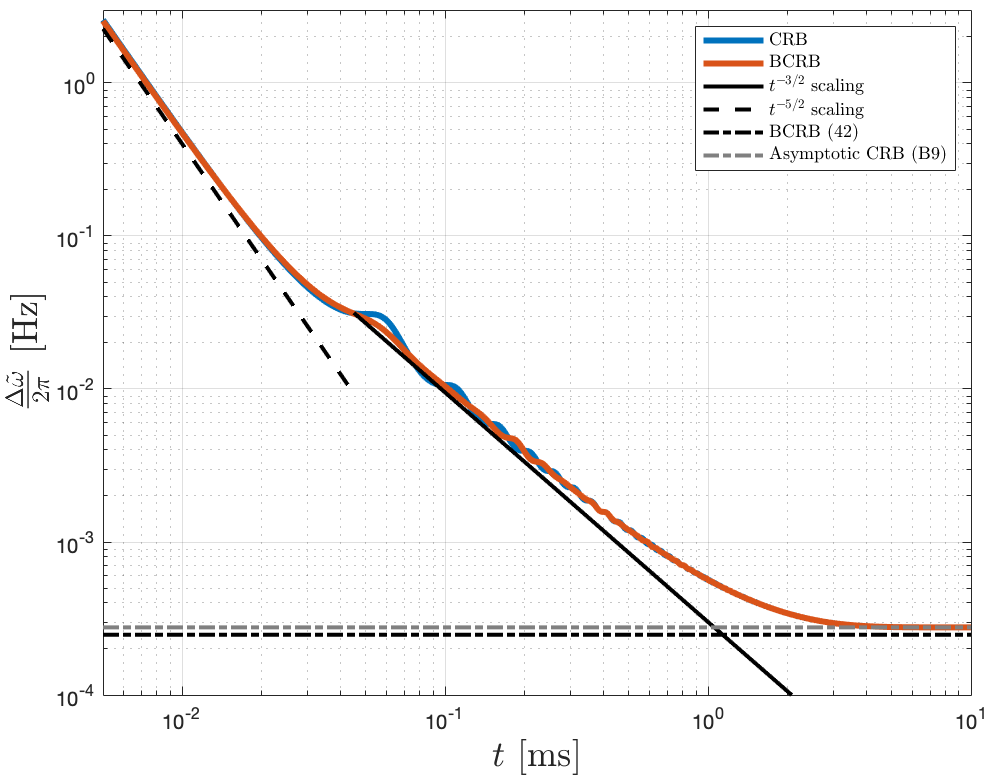}
    \end{center}
    \caption{\textbf{CRB \eqref{eq:CRB} and BCRB \eqref{eq:A6} as a function of time} computed numerically using the exact expressions for the relevant FI and its Gaussian prior average ($\sigma_\larmor=\SI{2}{\kilo\hertz}$), i.e., Eqs.~\eqref{eq:FI_no-noise},~\eqref{eq:A9} and~\eqref{eq:J_D_exact}. The dashed and solid black lines correspond to expressions \eqref{eq:FI_no-noise_short_time} and \eqref{eq:FI_no-noise_no-dec_larget}, respectively, predicting $t^{-5}$ and subsequent $t^{-3}$ scaling of the average MSE. The dash-dotted line marks the noiseless BCRB \eqref{eq:noiseless_bcrb} that is asymptotically (almost, recall approx.~\eqref{eq:JD_upper}) saturated. All parameters, apart from $Q=0$, are set as in \tabref{tab:exp_pars}.}
    \label{fig:exact_bcrb}
\end{figure}

For better illustration of the results, \figref{fig:exact_bcrb} shows CRB \eqref{eq:CRB} and BCRB \eqref{eq:A6} as functions of time, with the latter computed for a Gaussian prior with $\sigma_\larmor=2\SI{2}{\kilo \hertz}$, with all parameters set as in \tabref{tab:exp_pars}. The CRB and BCRB do not differ significantly---they both initially scales as $t^{-5}$ and subsequently as $t^{-3}$ in accordance with Eqs.~\eqref{eq:FI_no-noise_short_time} and \eqref{eq:FI_no-noise_no-dec_larget}, respectively, before ultimately saturating at (almost) the constant value predicted by \eqnref{eq:JD_upper}, equivalent then to the noiseless BCRB \eqref{eq:noiseless_bcrb}.

\section{Estimation of the number of atoms}
\label{app:natom_est}
We demonstrate here that the number of atoms, $N$, can be, in principle, directly determined from observations of the steady-state fluctuations of measurement outcomes $y_k$, i.e., when the atomic ensemble is reaches a completely depolarised state with $t\gg T_2$. 

According to the fluctuation-dissipation theorem~\cite{gardiner2004handbook} applied to the state-observation dynamics (\ref{eq:sys_PME}-\ref{eq:sys_output}) and given the dynamical matrices \eref{eq:SPM_dynamical_mats} applicable in the magnetometry setting of interest, the measurement outcomes $y_k$ \eqref{eq:SPM_obs_dyn} in the steady-state (thermal equilibrium) must follow a zero-mean Gaussian distribution with variance:
\begin{equation}
\Sigma_y=\frac{Ng_D^2q}{2}+\frac{R}{\Delta},
\label{eq:B1}
\end{equation}
where all the parameters stated above we assume to be perfectly known, e.g., taking values stated in \tabref{tab:exp_pars}.

Then, by replacing $\Sigma_y$ with its estimator, we obtain an unbiased estimate of the number of atoms:
\begin{equation}
\est{N}=\frac{2}{qg_D^2}\left(\est{\Sigma}_y-\frac{R}{\Delta}\right),
\label{eq:B2}
\end{equation}
where 
\begin{equation}
    \est{\Sigma}_y=\frac{1}{k-1}\sum\limits_{j=1}^ky_k^2,
    \label{eq:B3}
\end{equation}
is the variance estimator of $y_k$. 

Note that the average error of $\est{\Sigma}_y$ is given by
\begin{equation}
    \sqrt{\Delta^2\est{\Sigma}_y}=\sqrt{\frac{2}{k-1}}s^2=\sqrt{\frac{2}{k-1}}\left(\frac{Ng_D^2q}{2}+\frac{R}{\Delta}\right),
    \label{eq:B4}
\end{equation}
where the second equality follows from \eqnref{eq:B1}. Hence, it follows from \eqnsref{eq:B2}{eq:B4} that the average error of $\est{N}$ is given by:
\begin{equation}
\sigma_{\hat{N}}=\sqrt{\frac{2}{k-1}}\left(N+\frac{2R}{g_D^2q\Delta}\right).
\label{eq:B5}
\end{equation}

Thus, substituting for the parameters of \tabref{tab:exp_pars} and considering large $k=t/\Delta$, we obtain
\begin{equation}
\frac{\sigma_{\hat{N}}}{N}\approx\frac{200}{\sqrt{k}},
\label{eq:B6}
\end{equation}
which implies that the value of $N$ could be estimated up to $\sim10\%$ by tracking the measurement noise over $t\approx 4\cdot 10^6 \Delta=\SI{20}{\second}$, given the parameters of \tabref{tab:exp_pars}.

\bibliographystyle{myapsrev4-2}
\bibliography{BB_real}

\end{document}